# Observation of perfect Andreev reflection due to Klein paradox in a topological superconducting state

**Seunghun Lee[1,2], Valentin Stanev[1,2,3], Xiaohang Zhang[1,2], Drew Stasak[2], Jack Flowers[2], Joshua S. Higgins[1,4], Sheng Dai[5], Thomas Blum[6], Xiaoqing Pan[5,6], Victor M. Yakovenko[4,3,7], Johnpierre Paglione[1,4], Richard L. Greene[1,4], Victor Galitski[4,3,7], and Ichiro Takeuchi[2,1,4*]**

[1]Center for Nanophysics and Advanced Materials, University of Maryland, College Park, MD 20742, USA

[2]Department of Materials Science and Engineering, University of Maryland, College Park, MD 20742, USA

[3]Joint Quantum Institute, University of Maryland, College Park, MD 20742, USA

[4]Department of Physics, University of Maryland, College Park, MD 20742, USA

[5]Department of Chemical Engineering and Materials Science, University of California-Irvine, Irvine, CA 92697, USA

[6]Department of Physics and Astronomy, University of California-Irvine, Irvine, CA 92697, USA

[7]Condensed Matter Theory Center, University of Maryland, College Park, MD 20742, USA

*Corresponding author: takeuchi@umd.edu

In 1928, P. Dirac proposed a new wave equation to describe relativistic electrons[1]. Shortly afterwards, O. Klein solved a simple potential step problem for the Dirac equation and stumbled upon an apparent paradox – the potential becomes transparent when the height is larger than the electron energy. For massless particles, backscattering is completely forbidden in Klein tunneling, leading to perfect transmission through *any* potential barrier[2,3]. Recent advent of condensed matter systems with Dirac-like excitations, such as graphene and topological insulators (TIs), has opened the possibility of observing the Klein tunneling experimentally[4-6]. In the surface states of TIs, fermions are bound by spin-momentum locking, and are thus immune to backscattering due to time-reversal symmetry. Here we report the **observation of perfect Andreev reflection in point contact spectroscopy – a clear signature of Klein tunneling and a manifestation of the underlying "relativistic" physics of a proximity-induced superconducting state in a topological Kondo insulator.**

The extraordinary result of Klein's gedanken experiment illustrates the intrinsic connection between particles and anti-particles in relativistic quantum mechanics, and observing it ostensibly requires velocities close to the speed of light[2]. However, several condensed matter systems have recently appeared as unexpected platforms for study of relativistic effects. In materials like graphene and TIs, the Dirac equation emerges as an effective low-energy description of band electrons[4,5]. In graphene heterostructures, modulation of conductance as functions of electron trajectory and electrostatic potential profile has previously been used as a vehicle to probe Klein tunneling[5,7,8]. Here, we demonstrate an alternative way to directly observe Klein tunneling using a TI. In particular, we use point contact Andreev reflection (PCAR) measurements at the interface between a normal metal and a topological superconducting state (superconducting surface state of a TI). Perfect transmission of electrons through a finite barrier is manifested in the observed conductance doubling within the superconducting gap ($\Delta$). The conductance doubling is due to the charge, spin, and momentum conservation in the Andreev reflection process which requires a positively charged hole with opposite spin and momentum to be left behind[9-11]. In real experiments, however, the conductance enhancement is easily suppressed by various inevitable scattering mechanisms arising from non-ideal interface conditions, and the complete doubling of the conductance is very rarely observed. The extreme sensitivity to scattering makes Andreev reflection a unique tool suitable for detecting Klein tunneling. Spin-momentum locking of the Dirac states prohibits an incident electron normal to the interface from reflection irrespective of the microscopic details of the interface[12]. It results in complete absence of backscattering, and thus gives rise to topologically-protected perfect Andreev reflection manifested by exact doubling of the conductance. Such a direct probe for observing Dirac particles can lead to a better understanding of its condensed matter implementations and greater use of their properties in quantum transport devices.

To investigate how the presence of Dirac states at a TI surface affects the processes of particle transport governed by Andreev reflection, we use a PtIr tip to form a point-contact interface with a TI film in which superconductivity is induced through the proximity effect (Fig. 1a). We use heterostructures consisting of samarium hexaboride ($SmB_6$) and yttrium hexaboride ($YB_6$) to induce superconductivity in the Dirac surface states of $SmB_6$. $SmB_6$ is a topological Kondo insulator whose bulk gap at low temperatures ensures an insulating bulk sandwiched by topologically protected conducting surface layers[13-19]. This is a critical prerequisite for observing the effects originating solely from the topologically protected states[20]. The use of isostructural rare-earth-hexaboride superconductor $YB_6$ (with $T_c \approx 6.3$ K) as the layer underneath $SmB_6$ ensures that we have a pristine $SmB_6/YB_6$ interface fabricated by a sequential high-temperature growth necessary for achieving a robust proximity effect[21] (see the supplementary information (SI) Section 1.1, 1.2 and 2.1 for details).

As theoretically expected[6] and experimentally confirmed[22], the superconducting proximity effect occurring in such TI/superconductor heterostructures creates helical Cooper pairing on the surface of a TI. Due to the constraints imposed by the 2D surface states and the insulating bulk, incoming electrons with finite $p_z$ (*i.e.*, perpendicular to the surface) momenta do not participate in the transport at the interface between a normal metal and TI/superconductor heterostructure[23], since it is not possible to simultaneously match the energy and the momentum components parallel to the interface for both sides of the contact. Thus, the PtIr-$SmB_6/YB_6$ contact creates an interface where only in-plane transport (*i.e.*, momentum parallel to the plane of the surface states, $p_z = 0$) is allowed (Fig. 1a). It should be noted that the exclusive coupling to the surface states at the point contact transport to $SmB_6$ is in contrast with that to chalcogenide TI materials, where transport into the trivial bulk states provides alternative conduction channels, which completely mask the signatures of the topological states[20,22]. In addition, induced spin-momentum locking in a normal metal in contact with a TI has previously been observed as a result of the topological proximity effect[24-27]. Due to the spin-momentum locking on both sides, incident electrons are forbidden to reflect back (Fig. 1b). The perfect electron transmission to superconducting $SmB_6$ and concomitant hole generation result in the observed conductance doubling for the energy within the proximity-induced $\Delta$.

Point contact spectroscopy on $SmB_6/YB_6$ heterostructures measured at 2 K displays normalized differential-conductance ($dI/dV$) curves with doubling of the conductance within the bias voltage corresponding to the induced $\Delta$ for thicknesses of the $SmB_6$ layer in the 20-to-30 nm range. As seen in Fig. 1c and 1d, the observed doubling of the conductance is exact within the uncertainty due to the fitting procedure. In this regime, the $SmB_6$ layer is thick enough to have fully-developed topologically-protected surface states, and the superconducting proximity effect from the $YB_6$ can still be observed at the top surface (as depicted in the inset of Fig. 1a). For these $SmB_6/YB_6$ heterostructures, conductance doubling is consistently observed for point contacts made at different positions on the same sample surfaces (see SI Section 2.2). This attests to the robustness of conductance doubling (*i.e.*, perfect Andreev reflection) against the variation in contact due to microstructural change at different contact spots.

The best theory fit to the data, based on the Blonder, Tinkham and Klapwijk (BTK) theory[9] (see below), results in a proximity-induced $\Delta$ of $\approx 0.7$ meV, smaller than the bulk $\Delta$ of $YB_6$ ($\approx 1.3$ meV)[28,29], as expected. The temperature dependence and magnetic-field dependence of a conductance spectrum were measured on a separately fabricated $Au-SmB_6/YB_6$ structure (see SI Section 1.3 and 2.5, respectively), in which the junction is made with a Au thin film. The temperature dependence of $\Delta$ obtained from the $Au-SmB_6/YB_6$ structure shows the expected Bardeen–Cooper–Schrieffer (BCS) behavior (Fig. 1e) confirming that the enhancement in the $dI/dV$ spectrum is due to the proximity-induced superconductivity on the topologically-protected top surface of $SmB_6$.

The transmission and reflection of particles through an interface between a normal metal and a superconductor is described by the BTK theory[9]. In the BTK theory, a dimensionless parameter $Z$ is introduced to represent the interfacial barrier strength which reduces the transparency of the interface: perfect conductance doubling within $\Delta$ thus requires $Z \approx 0$ in the standard BTK theory. However, we show below that the superconductivity induced in the topologically protected surface states of $SmB_6$ inhibits electron reflection and cancels the effect of the barrier strength. Thus, even for finite $Z$ in $PtIr-SmB_6/YB_6$ point contact, there can be perfect electron transmission, directly discernable by the conductance doubling. Describing this phenomenon requires modification of the BTK theory to account for the role of the Dirac surface states of $SmB_6$ in the conductance spectrum (henceforth referred to as the Dirac-BTK theory).

The perfect Andreev reflection is a direct consequence of the presence of topologically protected surface states, as well as the absence of a bulk conduction channel. Based on our systematic study, when the thickness of the $SmB_6$ film is less than about 20 nm[21,30], the effect of the hybridization of the top and bottom surface states appears to become pronounced, opening a gap in the surface states dispersion and weakening the topological protection[22,31]. This accounts for the reduced conductance enhancement observed in the contact with the 10 nm $SmB_6/YB_6$ heterostructure at 2 K (Fig. 2b). To confirm the role of the robust bulk gap of $SmB_6$ in our observation, we have also performed PCAR measurement on $Sm_{1-x}Y_xB_6/YB_6$ heterostructures, where Sm is partially substituted by Y in the top layer to modify its electronic structure. Y ions are expected to act as donors increasing the electron carrier concentration in $SmB_6$, and thus generate conducting bulk states, which in turn give rise to transport channels not subjected to spin-momentum locking (see SI Section 1.4). Point contact spectra taken on $Sm_{0.8}Y_{0.2}B_6/YB_6$ and $Sm_{0.5}Y_{0.5}B_6/YB_6$ heterostructures at 2 K indeed show substantially reduced conductance enhancement ($\approx 1.5$) at zero bias (Fig. 2c and 2d).

When the surface of a $YB_6$ film is probed directly (with no $SmB_6$ layer on top), the point contact spectrum at 2 K yields an entirely different characteristic: the junction is now in the regime where tunneling has significant contribution, exhibiting reduced conductance in the gap region of $YB_6$ with a barrier strength ($Z \approx 1$) at the interface (extracted using the standard BTK model) (Fig. 2e). The gap value ($\Delta \approx 1.3$ meV) determined from the fit is consistent with the full $\Delta$ of $YB_6$[28,29]. In the other limit, when the thickness of $SmB_6$ is larger than 40 nm, the conductance spectrum at 2 K (Fig. 2f) does not show any feature that corresponds to proximity-induced superconductivity. Instead, the entire conductance spectrum shows Fano resonance – a familiar signature of the Kondo lattice physics of bulk $SmB_6$[19].

To illustrate the uniqueness of the perfect Andreev reflection observed here, we have surveyed the open literature on PCAR measurements performed on a variety of superconductors. Figure 3 plots zero-bias normalized $d\mathrm{I}/d\mathrm{V}$ (*i.e.*, conductance enhancement) versus $Z$ (obtained from the BTK fit) from 44 reports selected from 250 publications we have found on PCAR measurements (the list of publications and other details are provided in SI Section 2.3). The general trend is well captured by the standard BTK model (cyan line): any significant conductance enhancement is observed only when $Z < 0.3$. To the best of our knowledge, there have only been two reports in the literature where observed normalized conductance at zero bias is larger than 1.9. They are both Nb-Cu junctions: Soulen *et al.* obtained zero-bias normalized $d\mathrm{I}/d\mathrm{V} \approx 2$ with $Z \approx 0$[32], while Strijkers *et al.* observed zero-bias normalized $d\mathrm{I}/d\mathrm{V} \approx 1.9$ with $Z \approx 0.14$[33]. However, the conductance spectra from both reports display a distinct feature: substantial conductance dips immediately outside the conductance enhancement (see SI Section 2.6). This feature has been considered as a signature of Andreev reflection process affected by a superconducting proximitized layer in the normal metal side due to the highly transparent interface between Nb and Cu (*i.e.*, $Z \approx 0$). The absence of the substantial dip feature in our results thus indicates that perfect conductance doubling observed in the $PtIr-SmB_6/YB_6$ junction is of a different origin compared to those for the Nb-Cu contacts. In SI Section 2.6, we provide detailed explanation and comparison between our $PtIr-SmB_6/YB_6$ point contact spectra and the Nb-Cu point contact spectra reported in References 32 and 33.

There are many factors that cause scattering and thus contribute to the barrier strength $Z$ in the standard BTK theory[9,10]. In point contact spectroscopy experiments, it is often difficult to avoid the formation of an oxide layer at the surface. Even when the interface is formed in-situ under vacuum for thin film devices, the interfaces are defined as where the two disparate materials meet: the difference in the crystal structure and the atomic level surface microstructure including facets and terminations can lead to structural and compositional disorder and defects serving as scattering centers. Mechanical point contacts have the added complication due to local deformation of the tip. Furthermore, Fermi velocity mismatch also affects the reflection and transmission probabilities. Therefore, for almost all normal metal – topologically trivial superconductor junctions, $Z$ is finite, leading to the conductance enhancement significantly less than two as shown in Fig. 3. Figure 3 illustrates the unusual and distinct nature of the observed perfect conductance doubling for the $SmB_6$/$YB_6$ heterostructures with the thicknesses of the $SmB_6$ layer in the 20-30 nm range. According to the standard BTK theory, this implies $Z \approx 0$ for contacts to these particular heterostructures. However, we believe the actual materials-dictated factors determining $Z$ are all similar or identical for all heterostructures studied here (including ones with thin $SmB_6$ (10 nm) and Y-substituted $SmB_6$). Thus, we expect "materials-dictated $Z$" for junctions exhibiting perfect Andreev reflection to be at $\approx$ 0.4 which is the average extracted $Z$ values for the contacts with heterostructures without complete topological protection (navy pentagon in Fig. 3).

Now we turn to the mechanism of the perfect Andreev reflection for finite $Z$. To describe the transmission and the reflection processes at an interface between a normal metal and a superconducting TI (which in our case is a TI with proximity-induced superconductivity in the surface states), we modify the standard BTK theory[9], which describes the transport at a normal metal-conventional superconductor interface, by further considering the unique properties of a superconducting TI. The key in the modification is to take into account the interplay of the spin and the momentum of the electrons in the surface states of $SmB_6$ – a consequence of the non-trivial topology of the bulk band structure. These states are described by the Dirac Hamiltonian displaying spin-momentum locking (as manifested in helicity). As first shown by Klein, this can lead to perfect transmission through an arbitrarily large potential barrier: normal reflection of a Dirac particle requires a complete spin flip, and thus it is forbidden. The presence of such perfectly transmitting channels at the boundary between a topological material and a topological superconductor nullifies the effects of the boundary barrier (including the Fermi velocity mismatch), leading to perfect Andreev reflection (*i.e.*, the doubling of the conductance within $\Delta$ in a conductance spectrum)[12]. Bulk PtIr is a normal metal. However, the topological proximity effect can make PtIr in contact with $SmB_6$ topologically nontrivial, thereby satisfying the necessary condition for the perfect Andreev reflection [24-27]: the contact with the $SmB_6$ surface breaks the degeneracy of the two helicities in the PtIr tip, and in a region adjacent to the interface, only those states with the helicity matching the ones on the $SmB_6$ side exist. The strong spin-orbit coupling of PtIr itself can also play a role in this process[34,35] (see SI Section 3 for details).

Following the above discussion, we model the PtIr-$SmB_6$ boundary as a line dividing the normal and the superconducting regions in the plane of the $SmB_6$ surface states. On both sides, only states with the same helicity are allowed. At the boundary, we add a delta-function potential term $U(x) = U_0\delta(x)$ modeling a potential barrier at the interface, typically represented by the dimensionless barrier-strength parameter $Z$ ($Z \equiv U_0/\hbar v_F^S$). The Dirac Hamiltonian on the superconducting TI side can be written (in $\Psi = \left[\psi_{\uparrow,\varepsilon,\boldsymbol{p}}, \psi_{\downarrow,\varepsilon,\boldsymbol{p}}, \psi_{\uparrow,-\varepsilon,-\boldsymbol{p}}^*, \psi_{\downarrow,-\varepsilon,-\boldsymbol{p}}^*\right]$ basis) as[36]

$$H_{\text{hetero}} = \begin{bmatrix} v_F \boldsymbol{p} \cdot \boldsymbol{\sigma} - \sigma_0\mu + \sigma_0 U(x) & i\sigma_y\Delta \\ -i\sigma_y\Delta & v_F \boldsymbol{p} \cdot \boldsymbol{\sigma}^* + \sigma_0\mu - \sigma_0 U(x) \end{bmatrix}, \tag{1}$$

where $\boldsymbol{p}$ is momentum in the $x$-$y$ plane, $\mu$ is the chemical potential, $\boldsymbol{\sigma} \equiv \left[\sigma_x, \sigma_y\right]$ ($\left\{\sigma_0, \sigma_x, \sigma_y, \sigma_z\right\}$ is the set of the identity and the Pauli matrices in the spin space), and $v_F^S$ is the Fermi velocity on the $SmB_6$ side. ARPES measurements have found three Dirac cones at the $SmB_6$ surface[37,38], but we consider a simplified model with a single Dirac cone, which nevertheless retains the crucial feature of the surface states – their topological protection[38] (see SI Section 3 for details). $\Delta$ is the proximity-induced superconducting gap in the top surface of $SmB_6$ layer. This term arises from the spin-singlet $s$-wave superconductivity in $YB_6$, but its projection on the low-energy helical Dirac states is a mix of spin-singlet and spin–triplet states.

Following the BTK framework, we match the wave functions consisting of an incoming plane wave, and the transmitted and reflected solutions on each side of the boundary between the normal metal and the superconducting TI regions. Using the appropriate boundary condition for a metal with single helicity (see SI Section 3) and for energies close to the Fermi level, we can obtain the coefficients for each allowed process: $r_e$ – reflection as an electron; $r_h$ – Andreev reflection; $t_e$– transmission as an electron-like particle; $t_h$ – transmission as a hole-like particle. These coefficients depend on: 1) the energy (or bias voltage V), 2) $\theta_k$ - the angle of incidence measured from the normal to the boundary, 3) $Z$ which encodes the effects of the boundary barrier, and 4) $v_F^S/v_F^N$, Fermi velocity mismatch ($v_F^N$ is the Fermi velocity on the normal metal – PtIr – side).

The conductance ($G=d\mathrm{I}/dV$) through the interface is then given (at zero temperature) by:

$$G = \frac{d\mathrm{I}}{dV} = G_0 \int_{-\chi}^{\chi} (1 - |r_e(\theta_k)|^2 + |r_h(\theta_k)|^2) f_{\theta_k} \cos\theta_k \ d\theta_k \qquad (2)$$

where $f_{\theta_k}$ models the angular distribution of the incoming electrons, $\chi \equiv \arcsin(v_F^N/v_F^S)$, and $G_0$ is a constant. The angular dependence of the $r_e$ goes as $r_e(\theta_k)\sim \sin\theta_k$ – reflection as an electron at $\theta_k = 0$ requires a spin flip, which is forbidden by time-reversal symmetry (due to the fact that the overlap of the two spin states is zero) and thus $r_e(\theta_k = 0) = 0$ . Reproducing the observed perfect conductance doubling requires a rather narrow $f_{\theta_k}$ centered around $\theta_k = 0$ (see SI Section 3). In this quasi-one-dimensional case, there is perfect transmission irrespective of the barrier height and Fermi velocity mismatch – the essence of Klein tunneling (red line in Fig. 3). For $|eV| < \Delta$ this leads to $r_h(\theta_k \approx 0) = 1$, while for $|eV| \gg \Delta$ we have $r_h = 0$ – combining these two results with Eq. 2 immediately leads to conductance doubling: $G(|eV| < \Delta)/G(|eV| \gg \Delta) = 2$.

There have been numerous scanning tunneling microscopy (STM)[17,39,40], angle-resolved photoemission spectroscopy (ARPES)[37,38] and low-energy electron diffraction (LEED)[41] studies reporting observations of ubiquitous one-dimensional conducting atomic structures on the surfaces of $SmB_6$ due to reconstruction. Such structures provide one-dimensional channels necessary for the observation of perfect Andreev reflection (see SI Section 3 for details). Note that a two-dimensional topological insulator with inherently one-dimensional edge states can provide another platform to observe Klein tunneling in proximity with superconductors[42]. However, construction of device structures to exclusively probe Andreev reflection process in the edge states without parasitic bulk conduction contribution remains a challenge.

In summary, we have observed perfect Andreev reflection, a manifestation of Klein tunneling, using proximity-induced superconductivity in a 3D TI. Despite the formal similarity between Dirac excitations in graphene and TIs, there are important differences between the two with respect to Klein tunneling. In graphene, the degeneracy between sublattices of the honeycomb structure is crucial, whereas in TIs it is the time-reversal symmetry which directly prohibits backscattering. The unusual combination of the topologically protected surface states and the lack of the bulk states in thin layers of $SmB_6$ films has facilitated the observation of perfect Andreev reflection due to Klein tunneling. Perfect transmission renders transport of individual electrons across an interface dissipation-less, regardless of origins of the potential barrier and its variation, an attractive attribute for many device applications including quantum information processing[43] and high sensitivity detectors[44]. We envision Klein tunneling in TIs to be a platform for exploring a variety of novel interface transport phenomena including perfect spin-filters as governed by unadulterated spin-momentum locking[45].

## Acknowledgements

We acknowledge Yun Suk Eo for valuable discussions on properties of $SmB_6$, Frederick C. Wellstood for discussions on possible applications of superconducting Klein tunneling devices, and H. M. Iftekhar Jaim for assistance with X-ray measurements. This project was funded by ONR N00014-13-1-0635, ONR N00014-15-1-2222, AFOSR No. FA9550-14-10332, and by C-SPIN, one of six centers of STARnet, a Semiconductor Research Corporation program, sponsored by MARCO and DARPA. It was also supported by NSF (DMR-1410665). We acknowledge support from the Maryland NanoCenter. Victor Galitski was supported by DOE-BES (DESC0001911) and the Simons Foundation. This work was also supported in part by the Center for Spintronic Materials in Advanced infoRmation Technologies (SMART) one of centers in nCORE, a Semiconductor Research Corporation (SRC) program sponsored by NSF and NIST.

## Author Contributions

S.L., X.Z. and I.T. conceived the experiment. S.L. fabricated thin films and devices, and performed their characterization including point contact spectroscopy measurements with assistance from X.Z. and J.S.H. V.S., V.M.Y. and V.G. performed the theoretical calculations. D.S. analyzed the compositions of the films. J.F. performed literature survey on previous Andreev reflection experiments. S.D., T.B. and X.P. performed TEM measurements. V.M.Y., J.P., R.L.G. and V.G. helped with data interpretation and analysis and manuscript preparation. S.L., V.S., X.Z. and I.T. wrote the paper. I.T. supervised and coordinated the project. All authors discussed the results and commented on the manuscript.

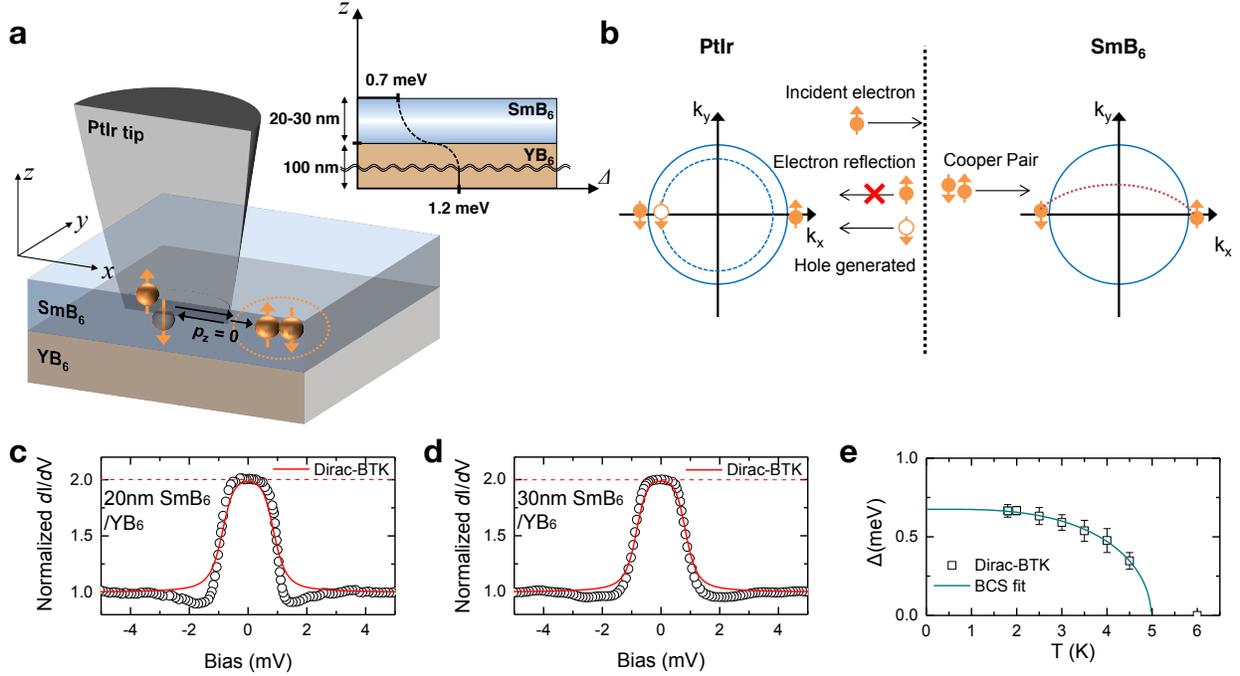

**Figure 1. Perfect Andreev reflection due to Klein paradox** (a) Schematic of point contact Andreev reflection (PCAR) measurement on $SmB_6$/$YB_6$ heterostructures. Due to the lack of bulk states in $SmB_6$, only electrons with momentum parallel to the plane of the surface states of $SmB_6$ (*i.e.*, $p_z = 0$) contribute to transport. The inset shows variation of the superconducting pair potential ($\Delta$) in $SmB_6$(20-30 nm)/$YB_6$ heterostructure. There is a finite $\Delta$ in the top conducting surface of $SmB_6$. (b) Andreev reflection process at the interface between PtIr and superconducting $SmB_6$. The surface of $SmB_6$ has topologically protected helical states exhibiting spin-momentum locking. Irrespective of barrier height in-between, normal electron reflection is not allowed as it requires a spin flip. Perfect Andreev reflection due to Klein tunneling, indicated by exact doubling of normalized differential-conductance ($dI$/$d$V), is observed in the point contact spectroscopy of (c) PtIr-$SmB_6$(20 nm)/$YB_6$(100 nm) and (d) PtIr-$SmB_6$(30 nm)/$YB_6$(100 nm) heterostructures measured at 2 K. The red lines are fits to the experimental data using a Blonder, Tinkham and Klapwijk model modified with Dirac Hamiltonian (Dirac-BTK, described in the text) with $\Delta$ = 0.75±0.06 meV for (c) and $\Delta$ = 0.73±0.05 for (d); (e) The temperature-dependent $\Delta$ (extracted using the Dirac-BTK model) from a Au-$SmB_6$ (20 nm)/$YB_6$ structure where Au thin film was used to form the junction (details in the supplementary information(SI) Section 1.3) displaying the Bardeen–Cooper–Schrieffer (BCS) behavior (cyan line).

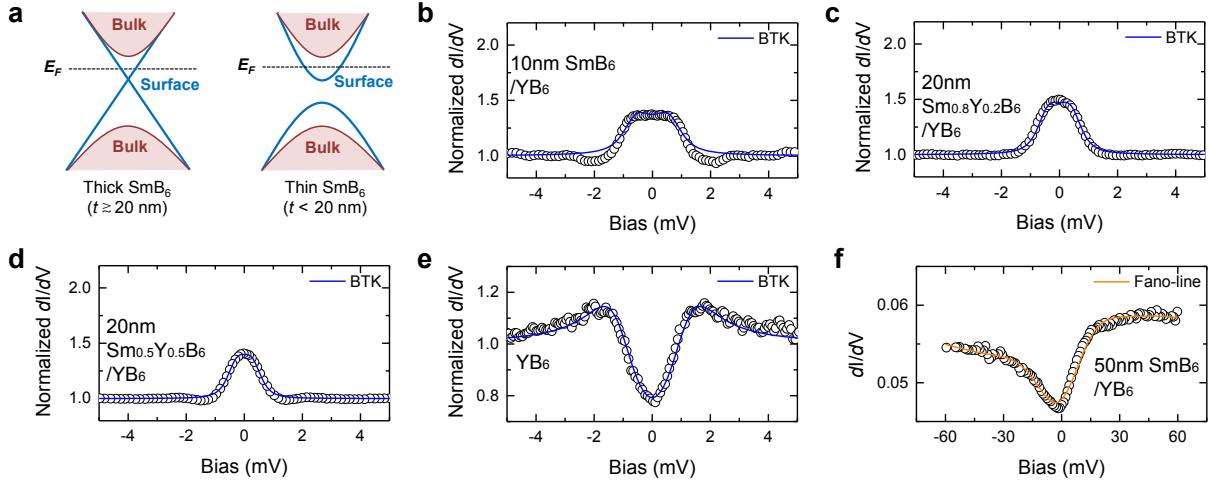

**Figure 2. Sensitivity of perfect Andreev reflection to compromised topological superconductivity:** when superconductivity in the surface state of $SmB_6$/$YB_6$ is modified by changing the thickness or the composition of the $SmB_6$ layer, the conductance doubling is suppressed. (a) Band structures of $SmB_6$ for different thicknesses: for thickness $t$ greater than 20 nm ($t \gtrsim 20$ nm, left) and for $t$ less than 20 nm ($t < 20$ nm, right); (b) Point contact spectrum of a 10 nm $SmB_6$/$YB_6$ heterostructure. Reduced conductance at zero bias (normalized $dI/dV \approx 1.4$) is observed. Point contact spectra of Y-substituted $SmB_6$ ($Sm_{1-x}Y_xB_6$)/$YB_6$ heterostructures with (c) x = 0.2 and (d) x = 0.5; and (e) point contact spectrum of $YB_6$ layer only. The blue lines are the best fits to the standard BTK theory (for (b), $Z = 0.42\pm0.10$, $\Delta = 0.59\pm0.10$ meV and $\Gamma \leqq 0.16$ meV; for (c), $Z = 0.35\pm0.09$, $\Delta = 0.49\pm0.05$ meV and $\Gamma \leqq 0.08$ meV; for (d), $Z = 0.42\pm0.06$, $\Delta = 0.30\pm0.04$ meV and $\Gamma \leqq 0.04$ meV; for (e), $Z = 1.04\pm0.06$, $\Delta = 1.24\pm0.08$ meV and $\Gamma = 0.60\pm0.04$ meV). $Z$ and $\Gamma$ are the interface barrier strength and the broadening parameter, respectively. (f) Point contact spectrum of a 50 nm $SmB_6$/$YB_6$ exhibits an asymmetric Fano-like spectrum due to the inherent Kondo-lattice electronic structure of $SmB_6$. The orange line is the best fit to the Fano-line shape [46]. All point contact spectra here are taken at 2 K.

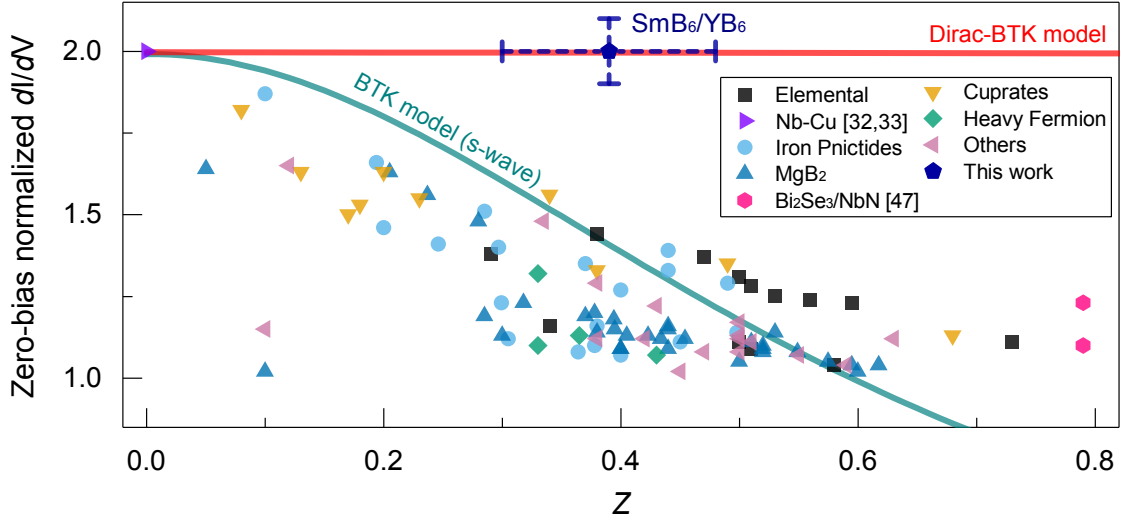

**Figure 3. Andreev reflection process under Dirac Hamiltonian** (a) A survey of reported zero-bias normalized $dI/dV$ (*i.e.*, conductance enhancement) versus $Z$ (*i.e.*, dimensionless barrier strength parameter) for PCAR measurements on a variety of superconductors (see SI Section 2.3 for references and plotted values). We note that there are several reports on point contact spectroscopy measurements of $Bi_2Se_3$/superconductor heterostructures[20,22,47], but only data from Ref. 47 meet the criteria for this figure and Table S1 as described in SI Section 2.3. The theoretical $Z$-dependent zero-bias normalized $dI/dV$ (*i.e.*, conductance enhancement) calculated by the standard BTK (cyan) and the Dirac-BTK models (red) are shown (described in text and in SI Section 2.4 and 3). For these curves, temperature of 2 K and $\Delta$=1 meV are used as the simulation parameters. For PtIr-$SmB_6$(20-30 nm)/$YB_6$ junctions displaying perfect Andreev reflection (normalized $dI/dV$ = 2), we assume the value $Z$ is similar to those for junctions with other heterostructures in this study without perfect Andreev reflection (*i.e.*, with thin $SmB_6$ (10 nm) and Y-substituted $SmB_6$): $Z$ = 0.39±0.09 with error bar reflecting the fitting procedure (navy pentagon, this work).

<Supplementary information>

# Observation of perfect Andreev reflection due to Klein paradox in a topological superconducting state

**1. Fabrication of SmB$_6$/YB$_6$ heterostructures and thin film devices**

    **1.1. Fabrication of SmB$_6$ thin films**

    **1.2. Fabrication of superconducting YB$_6$ thin films and the stoichiometry effect on *T$_c$***

    **1.3. Fabrication of Au-SmB$_6$/YB$_6$ structures and temperature-dependence of *d*I/*d*V curves**

    **1.4. Comparison of SmB$_6$ and Y-substituted SmB$_6$**

**2. Details of point contact spectroscopy measurements**

    **2.1. Measurement preparation**

    **2.2. Robustness of perfect Andreev reflection**

    **2.3. Conductance enhancement *vs. Z* barrier strength**

    **2.4. Comparison of conductance spectra in standard BTK and Dirac-BTK models**

    **2.5. Magnetic field-dependent conductance spectra of a point contact with a SmB$_6$/YB$_6$ heterostructure**

    **2.6. Conductance doubling and conductance dip near the gap**

**3. Details of the Modified Blonder, Tinkham and Klapwijk theory with the Dirac Hamiltonian**

# 1. Fabrication of SmB$_6$/YB$_6$ heterostructures and thin film devices

## 1.1. Fabrication of SmB$_6$ thin films

The growth conditions of SmB$_6$ thin films have been systematically optimized in order to ensure the quality of SmB$_6$ thin films. It is known that during the sputtering process, a significant difference in the atomic mass between Sm and B leads to different scattering probabilities, and thus likely results in a B-deficient film when the deposition is carried out with a stoichiometric target[1-3]. We, therefore, fabricated SmB$_6$ thin films on Si (001) substrates by co-sputtering SmB$_6$ and B targets to compensate possible B deficiency. To remove the native oxide layer on the Si substrate, we performed hydrofluoric acid (HF) treatment prior to the thin film deposition. After reaching a base pressure of ≈ 2×10$^{-8}$ Torr, the sputtering process was performed on the Si substrates at 860 ℃ under a deposition pressure of 10 mTorr of Ar (99.999 %). Distance between the targets and substrates as well as plasma density were adjusted to increase the activation energy of sputtered species which is correlated with chemical reaction and atomic migration[4]. We optimized the power ratio of the two targets for the co-sputtering process by measuring the stoichiometry (*i.e.*, B/Sm ratio) of the deposited SmB$_6$ thin films using wavelength dispersive spectroscopy (WDS). The optimal powers for SmB$_6$ and B were found to be 40W and 60W, respectively, for the distance between the targets and the substrate of ≈ 10 cm. Under the optimized condition, the B/Sm ratio of the SmB$_6$ thin film was 6.0±0.1. X-ray photoemission spectroscopy (XPS) and energy-dispersive spectroscopy (EDS) measurements of the films were used to verify the absence of any impurities which may give rise to metallic conduction at low temperatures. Temperature-dependent resistance measurements show the suggested signature of the emergence of metallic surface states – the saturation of the resistance at low temperatures (*i.e.*, resistance plateau) (see Section 1.4).

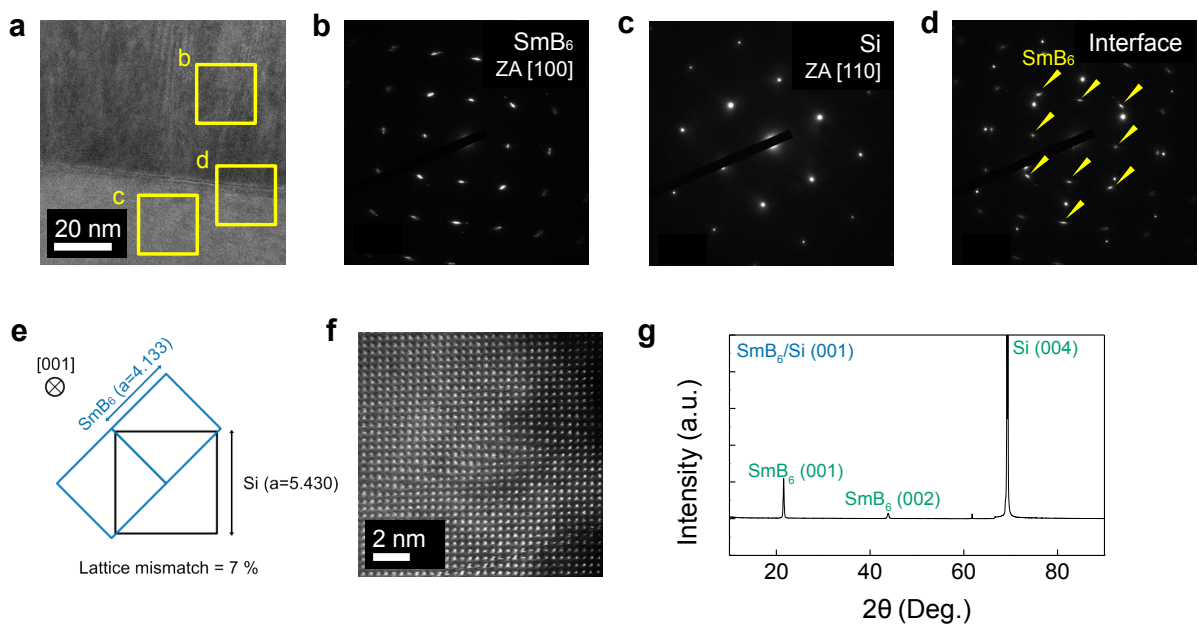

**Figure S1.** (a) High-resolution cross-sectional transmission electron microscopy image of a SmB$_6$ thin film. The yellow squares herein correspond to the regions for selected area electron diffraction (SAED) measurements shown in (b)-(d), *i.e.*, SmB$_6$, Si substrate, and SmB$_6$/Si interface regions, respectively (ZA: zone axis). (e) Epitaxial relationship between SmB$_6$ and Si substrate. (f) Aberration-corrected scanning transmission electron microscopy cross-sectional image of a SmB$_6$ thin film. (g) θ−2θ X-ray diffraction pattern of a SmB$_6$ thin film on a Si (001) substrate.

Figure S1a shows a high-resolution transmission electron microscopy image of a cross-section SmB$_6$ sample. There is no indication of the presence of interfacial gradation or extra phases. Figures S1b-d show selected area electron diffraction (SAED) patterns of the SmB$_6$ thin film, the Si substrate, and the interface regions, respectively. The SAED pattern of the Si substrate (Figure S1c) shows the pattern along the [110] zone axis (ZA). In the SAED pattern of the interface (Fig. S1d), an additional spot pattern corresponding to the SmB$_6$ [100] zone orientation (Fig. S1b) can be clearly identified (indicated by yellow arrows). The result is indicative of the epitaxial relation, SmB$_6$ [100] || Si [110], which is consistent with a small lattice mismatch between Si (110) and SmB$_6$ (100) as illustrated in Fig. S1e. Specifically, the *d*-spacing of Si (110) is 3.839 Å, and the lattice mismatch between Si (110) and SmB$_6$ (100) is about 7 %. In addition, aberration-corrected scanning transmission electron microscopy was utilized, and the atomic-resolution image taken from the SmB$_6$ film (Fig. S1f) displays its cubic structure. The θ−2θ X-ray diffraction pattern (Fig. S1g) shows a c-axis-oriented structure of SmB$_6$. The XRD diffraction pattern exhibits sharp SmB$_6$ peaks, which are associated with the {001} planes only. The lattice parameter is found to be 4.13 Å, which is close to the bulk value[5].

## 1.2. Fabrication of superconducting YB$_6$ thin films and the stoichiometry effect on $T_c$

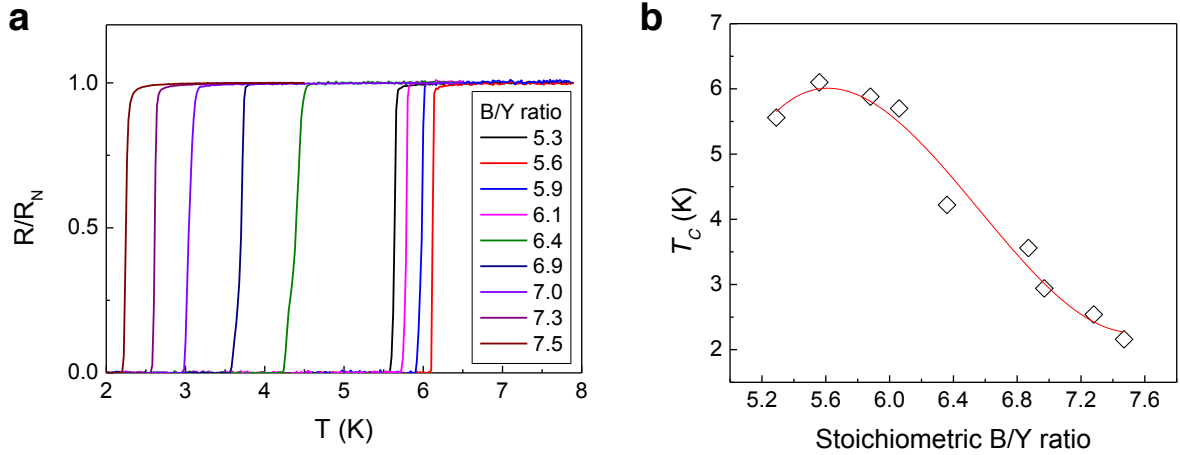

**Figure S2.** (a) Temperature-dependent resistance curves of YB$_{6\pm x}$ thin film for different stoichiometric B/Y ratios. (b) Change in $T_c$ as a function of stoichiometric B/Y ratio.

Yttrium hexaboride (YB$_6$) is a known rare-earth hexaboride superconductor with a bulk zero resistance $T_c \approx 7$ K[6,7]. It has been reported that the superconducting properties of YB$_6$ are closely related to the composition[8]. However, systematic study of the superconducting properties in a broad range of compositional variation has not been previously reported. We have successfully fabricated superconducting YB$_{6\pm x}$ films for the first time. To achieve the highest $T_c$ in YB$_{6\pm x}$ thin films for the present study, we first studied the stoichiometry effect on the $T_c$ of sputtered YB$_{6\pm x}$ thin films. Due to the substantial difference in the atomic mass between Y and B, and the variation in the distance from the target to different locations of a 3" wafer, we were able to fabricate "natural composition spread" films of YB$_{6\pm x}$ by sputtering a stoichiometric YB$_6$ target. Similar to the deposition and the characterization of SmB$_6$ thin films, a deposition pressure of 10 mTorr and a growth temperature of 860 ˚C were used for YB$_{6\pm x}$ thin film growth on Si (001) substrates. The distance between the YB$_6$ target and the Si substrate was ≈ 10 cm, and the DC power applied to the YB$_6$ target was 60 W. The stoichiometric B/Y ratio for films deposited at different positions was examined by WDS measurements. As shown in Figure S2a, the temperature dependence of the normalized resistance (R/R$_N$ (R$_N$: normal state resistance)) of the YB$_{6\pm x}$ thin films indicates that the superconducting transition temperature, $T_c$, varies with the stoichiometric B/Y ratio. In Fig. S2b, $T_c$ is plotted as a function of the stoichiometric B/Y ratio. The highest $T_c$ is observed in the slightly boron deficient region (B/Y = 5.6). Thus, YB$_{5.6}$ films were used for the present study, and for simplicity, the YB$_{5.6}$ films used in this study are referred to as YB$_6$ films. The SmB$_6$/YB$_6$ heterostructures were fabricated through a sequential high-temperature deposition process without breaking the vacuum, *i.e.*, an in-situ process as described in the main text, to ensure a pristine interface between SmB$_6$ and YB$_{5.6}$[2]. YB$_6$ has a cubic structure with almost the same lattice constant as SmB$_6$ ($\approx$ 4.1 Å) (YB$_6$: JCPDS no. 16-0732 and SmB$_6$: JCPDS no. 36-1326), and thus lattice mismatch strain is expected to be negligible.

### 1.3. Fabrication of Au-SmB$_6$/YB$_6$ structures and temperature-dependence of $dI/dV$ curves

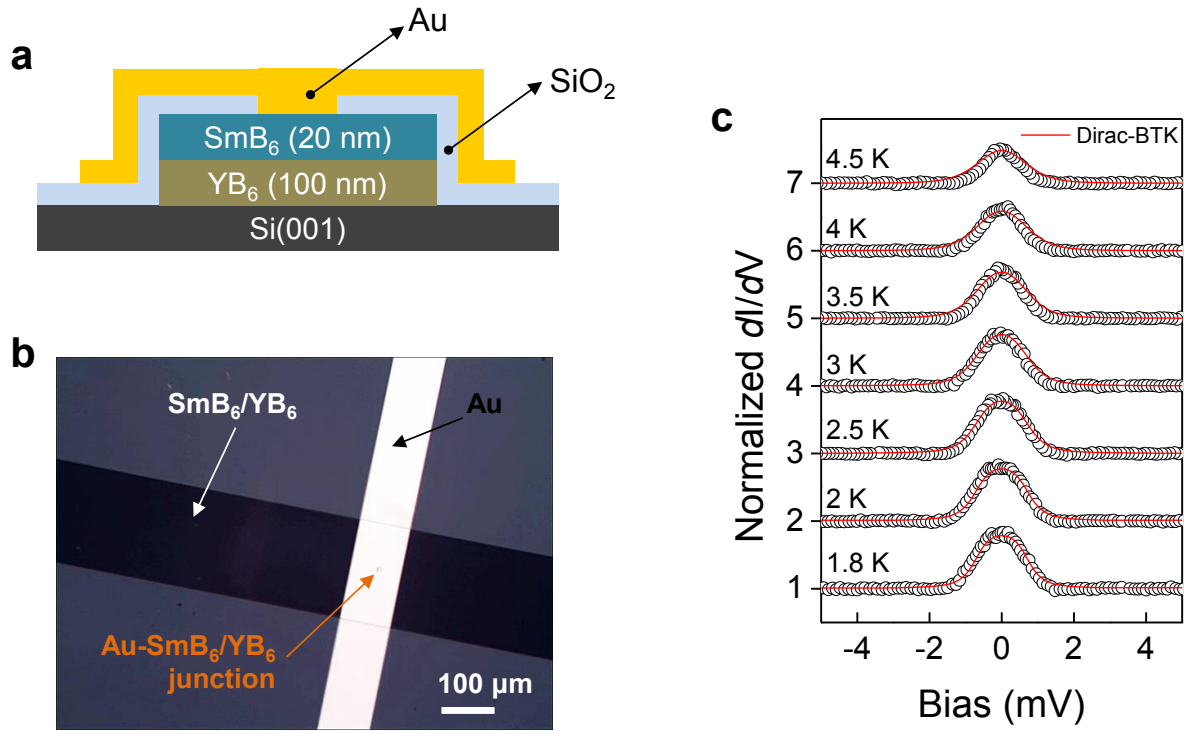

**Figure S3.** (a) Cross-sectional schematic of a Au-SmB$_6$(20nm)/YB$_6$(100nm) structure. (b) Optical microscope image of the device. (c) Normalized $dI/dV$ spectra of the Au-SmB$_6$/YB$_6$ structure at different temperatures. The red lines are fits using the Dirac-BTK model (see the main text or Section 3 in the supplementary information). The normalized $dI/dV$ at 1.8 K are plotted with the actual value, while other curves are vertically shifted for clarity.

The analysis of temperature-dependence of $dI/dV$ spectra can be used to verify that the gap-like feature is indeed attributed to the proximity-induced superconductivity. To perform a systematic temperature-dependence measurement, Au-SmB$_6$/YB$_6$ structures were fabricated using a method including multiple photolithography and ion milling processes. Microposit S1813 was used as the photoresist, and after spin coating, the photoresist was baked at 100 ℃ for two minutes. After Ultra-violet (UV) light exposure, the samples were developed using a Microposit CD-26 developer for 60 seconds. A schematic cross-sectional structure of the thin film devices is shown in Figure S3a. In-situ Ar plasma cleaning was performed on SmB$_6$/YB$_6$ heterostructure prior to Au deposition. After the Au deposition, two rounds of photolithography and ion milling processes were carried out to define the line shape and the circular junction area, respectively. SiO$_2$ (100 nm) layer was used to electrically isolate a top electrode from the SmB$_6$/YB$_6$ line. The top electrode, consisting of Au, was fabricated through a lift-off process. The optical microscope image (Figure S3b) shows the top view of a Au-SmB$_6$/YB$_6$ structure with a circular junction (diameter = 10 μm).

Figure S3c shows normalized $d$I/$d$V spectra of the Au-SmB$_6$(20 nm)/YB$_6$ structure at different temperatures (1.8 – 4.5 K). The zero-bias conductance enhancement due to Andreev reflection is $\sim$ 1.8 which is slightly smaller than the value obtained from junctions in the point-contact configuration as described in the main text. Given the specific geometric design of the junction, quasiparticle lift-time broadening[9,10] and/or an oblique angle for incident electrons may lead to the slightly reduced zero-bias conductance enhancement (see Section 3). In the former case, for example, as shown in Fig. S3c, by introducing a life-time broadening term $\Gamma$ with a value of less than 10 % of $\Delta$, we can fit the data using the Dirac-BTK model (the Blonder, Tinkham and Klapwijk model modified with the Dirac Hamiltonian). The $\Delta$ values obtained by the Dirac-BTK fits on the $d$I/$d$V spectra at different temperatures agree well with those from point contact spectroscopy measurements carried out with a PtIr tip.

### 1.4. Comparison of SmB$_6$ and Y-substituted SmB$_6$

To confirm that the absence of bulk gapless states is crucial for the perfect conductance doubling observed in point contact spectroscopy measurements, we modified the bulk electronic structure of SmB$_6$ by Y substitution. Specifically, we performed point contact spectroscopy measurements on Sm$_{1-x}$Y$_x$B$_6$/YB$_6$ heterostructures. Y-substituted SmB$_6$ were prepared by co-sputtering SmB$_6$, B, and YB$_6$ targets, and the composition was determined by WDS. Figure S4a shows the resistance normalized by the value at 300 K ($R/R_{300K}$, log scale) *vs.*

the inverse of temperature ($1/T$) plots of SmB$_6$ as well as 20% and 50% Y-substituted SmB$_6$ (Sm$_{0.8}$Y$_{0.2}$B$_6$ and Sm$_{0.5}$Y$_{0.5}$B$_6$) thin films. The behavior of temperature-dependent resistance of the bulk states can be described by an exponential function, $R_{bulk}(T) \propto \exp(E_a/k_B T)$, where $E_a$ and $k_B$ are a carrier activation energy and Boltzmann constant, respectively. Hence, the positive linear slopes in the relatively high-temperature region in Fig. S4a are approximately proportional to the corresponding activation energies. The slope decreases with increasing the Y concentration, which implies that Y-substitution increases the bulk conductivity and reduces the activation energy of carriers. More explicitly, in order to estimate and provide the activation energies of SmB$_6$ and Sm$_{0.8}$Y$_{0.2}$B$_6$, only bulk conductance channel should be taken into account. Thus, based on a simple parallel conductance model (total $G = G_{bulk} + G_{surface}$) below the temperature where the Kondo gap is completely open ($\approx$ 40 K)[2,3,11,12], we plot $G$ - $G_{surface}$ (log scale) $vs.$ $1/T$ in Fig. S4b where $G_{surface}$ is modeled as a linear function of temperature[11], and $G$ - $G_{surface}$ are normalized by $G$ at 300 K. Now the slopes of $G$ - $G_{surface}$ in this figure correspond to the activation energies of pure SmB$_6$ and Sm$_{0.8}$Y$_{0.2}$B$_6$ which are found to be 3.0 meV and 2.2 meV, respectively.

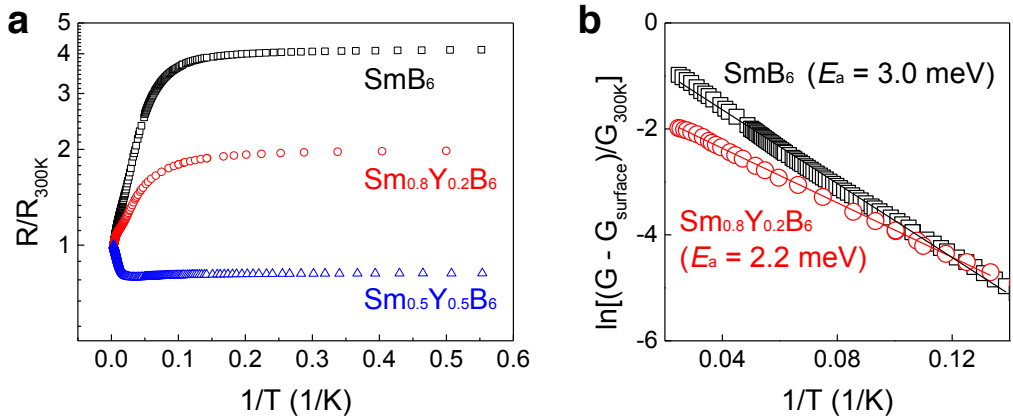

**Figure S4.** (a) Comparison of log $R$ $vs.$ $1/T$ behaviors of SmB$_6$, and 20 % and 50 % Y-substituted SmB$_6$ (*i.e.*, Sm$_{0.8}$Y$_{0.2}$B$_6$ and Sm$_{0.5}$Y$_{0.5}$B$_6$, respectively). The resistance values are normalized by their values at 300 K. The positive linear slopes at relatively high-temperature region are roughly proportional to the activation energy. (b) $G$ - $G_{surface}$ (log scale, normalized by conductance $G$ at 300 K) $vs.$ $1/T$ of pure SmB$_6$ (black squares) and Sm$_{0.8}$Y$_{0.2}$B$_6$ (red circles). Each slope of the linear fits (black and red lines) corresponds to the activation energy ($E_a$) of pure SmB$_6$ and Sm$_{0.8}$Y$_{0.2}$B$_6$, which are 3.0 meV and 2.2 meV, respectively.

## 2. Details of the point contact spectroscopy measurements

### 2.1. Measurement preparation

Point contact Andreev reflection measurements were carried out using a home-built probe designed for operation in a Physical Property Measurement System (Quantum Design, Inc.). Using a mechanically sharpened tip, the point contact junctions with contact resistance of few ohms were achieved by gently approaching the tip onto the surface of the heterostructure at 2 K.

### 2.2. Robustness of perfect Andreev reflection

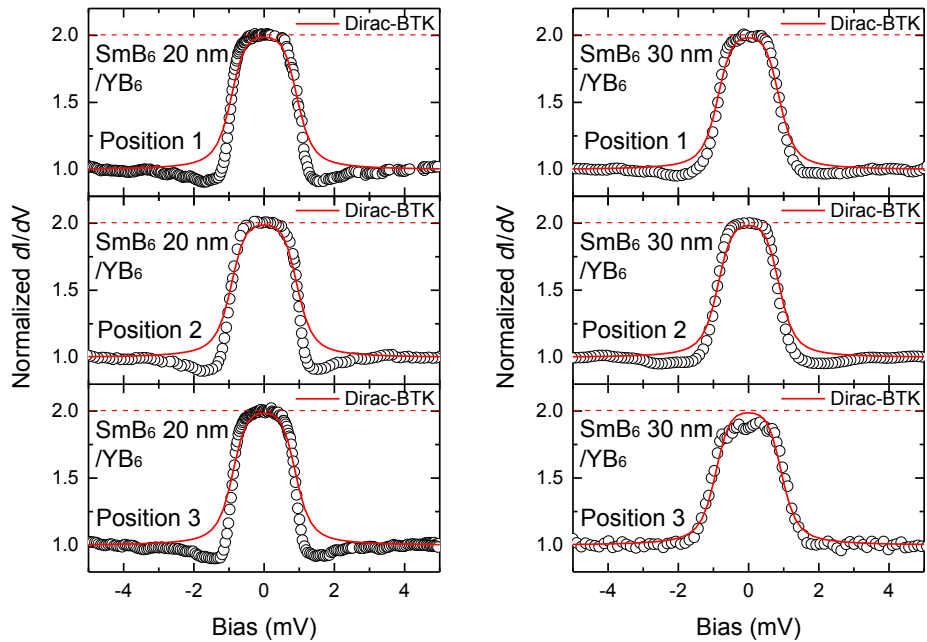

**Figure S5.** Point contract spectra obtained at different positions (Position 1, Position 2, and Position 3, which are roughly 1 mm apart from each other) on $SmB_6/YB_6$ heterostructures with 20 nm thick $SmB_6$ (left panel) and 30 nm thick $SmB_6$ (right panel). Conductance doubling is consistently observed at all position in the conductance spectra of the $SmB_6/YB_6$ heterostructures.

In order to demonstrate the robustness of perfect Andreev reflection observed in the $SmB_6(20\text{-}30 \text{ nm})/YB_6$ heterostructures, we have made multiple contact measurements by lifting up the PtIr tip and repositioning it to land at other spots (Position 1-3) on the same samples. As shown in Fig. S5, in each set of such measurements, we have consistently obtained conductance doubling for all contacts made on $SmB_6(20\text{-}30 \text{ nm})/YB_6$ heterostructures despite the expected local variation in the surface microstructure.

## 2.3. Conductance enhancement *vs. Z* barrier strength

To illustrate the uniqueness of the perfect Andreev reflection that is evident here in the doubled conductance (zero-bias normalized $dI/dV$ = 2) and the difficulty in general in observing such a high conductance enhancement, we have surveyed the open literature on point contact spectroscopy measurements on a variety of superconductors. Because $Z$ is a primary parameter associated with the conductance enhancement in the standard BTK theory[13], we plot zero-bias normalized $dI/dV$ *vs. Z* in Fig. 3 in the main text. We have looked at over about 250 publications on point contact spectroscopy measurements and selected data points from 44 reports using two criteria; 1) the value of $Z$ is extracted using a BTK fit, and 2) the conductance enhancement is larger than one (normalized $dI/dV \geq 1$), which indicates that a particular junction is not in the tunnel-dominant regime, and it is not governed by any zero-bias conductance peak due to a nodal order parameter. For the plot (Fig. 3 in the main text), we display the data points in the range of $0 \leq Z \leq 0.8$. The detailed information including the types of superconductors, contacts, and their references are summarized in Table S1.

**Table S1.** List of the reported zero-bias normalized $dI/dV$ (i.e., conductance enhancement) versus $Z$ values from different point contact Andreev reflection measurements (plotted in Fig. 3 in the main text)

| Superconductor | Metal contact | Conductance enhancement | Z-value | Reference |
|---|---|---|---|---|
| $AuIn_2$ | Cu tip | 1.08 | 0.50 | Gloos and Martin, *Z. Phys. B* **99**, 321 (1996) |
| $AuIn_2$ | Cu tip | 1.15 | 0.10 | |
| $Ba_{0.23}K_{0.77}Fe_2As_2$ | Pb tip | 1.87 | 0.10 | Zhang *et. al., Phys. Rev. B* **82**, 020515 (2010) |
| $Ba_{0.55}K_{0.45}Fe_2As_2$ | Pt tip | 1.11 | 0.45 | Samuely *et. al., Physica C* **469**, 507 (2009) |
| $CeCoIn_5$ | Au tip | 1.13 | 0.37 | Park *et. al., Phys. Rev. B.* **72**, 052509 (2005) |
| $Cu_xBi_2Se_3$ | Pd/Cr electrode | 1.02 | 0.45 | Peng *et al., Phys. Rev. B* **88**, 024515 (2013) |
| $Cu_xBi_2Se_3$ | Pd/Cr electrode | 1.12 | 0.38 | |

| Superconductor | Metal contact | Conductance enhancement | Z-value | Reference |
|---|---|---|---|---|
| $Bi_2Se_3$/NbN | Au | 1.10 | 0.79 | Koren and Kirzhner, *Phys. Rev. B* **86**, 144508 (2012)) |
| $Bi_2Se_3$/NbN | Au | 1.23 | 0.79 | |
| $La_{1.85}Sr_{0.15}CuO_4$ | Au tip | 1.82 | 0.08 | Gonnelli *et. al., Eur. Phys. J. B* **22**, 411 (2001) |
| $La_{1.87}Sr_{0.13}CuO_4$ | Au tip | 1.50 | 0.17 | |
| $La_{1.88}Sr_{0.12}CuO_4$ | Au tip | 1.53 | 0.18 | |
| $La_{1.8}Sr_{0.2}CuO_4$ | Au tip | 1.63 | 0.13 | |
| $La_{1.92}Sr_{0.08}CuO_4$ | Au tip | 1.63 | 0.20 | |
| $La_{1.9}Sr_{0.1}CuO_4$ | Au tip | 1.55 | 0.23 | |
| $LaFeAsO_{0.9}F_{0.1}$ | Ag paint | 1.14 | 0.50 | Gonnelli *et. al, Physica C* **469**, 512 (2009) |
| $LaFeAsO_{0.9}F_{0.1}$ | Ag paint | 1.08 | 0.36 | |
| $LaFeAsO_{0.9}F_{0.1}$ | Ag paint | 1.07 | 0.40 | Gonnelli *et. al., Phys. Rev. B* **79**, 184526 (2009) |
| $LiTi_2O_4$ | PtIr tip | 1.65 | 0.12 | Tang *et. al., Phys. Rev. B* **73**, 184521 (2006) |
| $LuNi_2B_2C$ | Cu tip | 1.08 | 0.50 | Kvitnitskaya *et. al., Supercond. Sci. Technol.* **23**, 115001 (2010) |
| $Mg(B_{0.868}C_{0.132})_2$ | Ag paint | 1.10 | 0.52 | Daghero *et. al., Phys. Stat. Sol. C* **2**, 1656 (2005) |
| $Mg(B_{0.895}C_{0.105})_2$ | Ag paint | 1.20 | 0.38 | |
| $Mg(B_{0.907}C_{0.093})_2$ | Ag paint | 1.12 | 0.43 | |
| $Mg(B_{0.917}C_{0.083})_2$ | Ag paint | 1.08 | 0.55 | |
| $Mg(B_{0.945}C_{0.055})_2$ | Ag paint | 1.11 | 0.51 | |
| $Mg(B_{0.953}C_{0.047})_2$ | Ag paint | 1.18 | 0.39 | |

| Superconductor | Metal contact | Conductance enhancement | Z-value | Reference |
|---|---|---|---|---|
| $Mg(B_{0.9}C_{0.1})_2$ | Ag paint | 1.12 | 0.45 | Daghero *et. al., Phys. Stat. Sol. C* **2**, 1656 (2005) |
| $Mg_{0.6}(AlLi)_{0.4}B_2$ | Ag paint | 1.15 | 0.40 | Daghero *et. al., Supercond. Sci. Technol.* **22**, 025012 (2009) |
| $Mg_{0.6}(AlLi)_{0.4}B_2$ | Ag paint | 1.16 | 0.44 | |
| $Mg_{0.68}Al_{0.32}B_2$ | Ag paint/Au tip | 1.09 | 0.40 | Gonnelli *et. al., J. Supercond. Nov. Magn.* **20**, 555 (2007) |
| $Mg_{0.7}(AlLi)_{0.3}B_2$ | Ag paint | 1.13 | 0.41 | Daghero *et. al., Supercond. Sci. Technol.* **22**, 025012 (2009) |
| $Mg_{0.7}(AlLi)_{0.3}B_2$ | Ag paint | 1.05 | 0.58 | |
| $Mg_{0.8}(AlLi)_{0.2}B_2$ | Ag paint | 1.09 | 0.44 | |
| $Mg_{0.82}Al_{0.18}B_2$ | Ag paint | 1.13 | 0.42 | Daghero *et. al., Phys. Stat. Sol. C* **2**, 1656 (2005) |
| $Mg_{0.9}(AlLi)_{0.1}B_2$ | Ag paint | 1.23 | 0.32 | Daghero *et. al., Supercond. Sci. Technol.* **22**, 025012 (2009) |
| $Mg_{0.91}Al_{0.09}B_2$ | Ag paint | 1.09 | 0.40 | Daghero *et. al., Phys. Stat. Sol. C* **2**, 1656 (2005) |
| $Mg_{0.92}Al_{0.08}B_2$ | Ag paint | 1.08 | 0.52 | |
| $Mg_{0.98}Al_{0.02}B_2$ | Ag paint | 1.08 | 0.52 | |
| $MgB_2$ | Pt tip | 1.63 | 0.21 | Gonnelli *et. al., J. Phys. Chem. Solids* **63**, 2319 (2002) |
| $MgB_2$ | Pt tip | 1.56 | 0.24 | |
| $MgB_2$ | Pt tip | 1.19 | 0.37 | |
| $MgB_2$ | Au tip | 1.14 | 0.38 | |
| $MgB_2$ | Cu tip | 1.09 | 0.52 | Szabo *et. al, Phys. Rev. Lett.* **87**, 137005 (2001) |
| $MgB_2$ | Pt tip | 1.05 | 0.50 | Laube *et. al., EPL* **56**, 296 (2001) |
| $MgB_2$ | Pt tip | 1.13 | 0.30 | Laube *et. al., EPL* **56**, 296 (2001) |

| Superconductor | Metal contact | Conductance enhancement | Z-value | Reference |
|---|---|---|---|---|
| $MgB_2$ | Pt tip | 1.02 | 0.10 | |
| $MgB_2$ | Au tip | 1.14 | 0.53 | Bugoslavsky *et. al., Supercond. Sci. Technol.* **15**, 526 (2002) |
| $MgB_2$ | PtIr tip | 1.19 | 0.29 | |
| $MgB_2$ | PtIr tip | 1.48 | 0.28 | Li *et. al., Phys. Rev. B* **66**, 064513 (2002) |
| $MgB_2$ | PtIr tip | 1.64 | 0.05 | |
| $MgB_2$ | In spot | 1.02 | 0.60 | Gonnelli *et. al., Phys. Rev. Lett.*, **89**, 247004 (2002) |
| $MgB_2$ | Ag paint or In spot | 1.04 | 0.60 | Daghero et. al. *Physica C* **385**, 255 (2003) |
| $MgB_2$ | Ag paint | 1.15 | 0.44 | Gonnelli *et. al., Phys. Rev.* B, **69** 100504 (2004) |
| $MgB_2$ | Ag paint | 1.04 | 0.62 | Daghero *et. al., Phys. Stat. Sol. C* **2**, 1656 (2005) |
| $MgCNi_3$ | PtIr tip | 1.48 | 0.34 | Shan *et. al., Phys. Rev. B* **68**, 144510 (2003) |
| Nb | Ag electrode | 1.04 | 0.58 | Naidyuk *et. al., Physica B* **218**, 122 (1996) |
| Nb | Cu | 1.28 | 0.51 | Kant *et al, Phys. Rev. B* **66**, 212403 (2002) |
| Nb | Ag tip | 1.31 | 0.50 | |
| Nb | Ag tip | 1.37 | 0.47 | Wei *et. al., Appl. Phys. Lett.* **97**, 062507 (2010) |
| Nb | Ag tip | 1.11 | 0.73 | |
| Nb | Cu | 1.92 | 0.14 | Strijkers *et al., Phys. Rev. B* **63**, 104510 (2001) |
| Nb | Cu | 2.00 | 0.00 | Soulen *et al., Science* **282**, 85 (1998) |
| Nb/Cu(65% Nb) | PtIr tip | 1.29 | 0.38 | Parab *et. al., Supercond. Sci. Technol.* **30**, 055005 (2017) |
| Nb/Cu(87% Nb) | PtIr tip | 1.04 | 0.59 | Parab *et. al., Supercond. Sci. Technol.* **30**, 055005 (2017) |

| Superconductor | Metal contact | Conductance enhancement | Z-value | Reference |
|---|---|---|---|---|
| NdFeAsO$_{0.85}$ | Au tip | 1.39 | 0.44 | Yates *et. al.,* Supercond. Sci. Technol. **21**, 092003 (2008) |
| Pb | Al | 1.25 | 0.53 | Kant *et. al,* Phys. Rev. B **66**, 212403 (2002) |
| Pb | Cu nanocontact | 1.38 | 0.29 | Chalsani *et. al.,* Phys. Rev. B **75**, 094417 (2007) |
| Pb | Cu/Pt nanocontact | 1.16 | 0.34 | |
| Pr$_{1.83}$Ce$_{0.17}$CuO$_4$ | Au or PtRh contact | 1.33 | 0.38 | Qazilbash *et. al.,* Phys. Rev. B, **68**, 024502 (2003) |
| SmFeAsO$_{0.75}$F$_{0.25}$ | Cu tip | 1.16 | 0.38 | Naidyuk *et. al.,* Supercond. Sci. Technol. **24**, 065010 (2011) |
| SmFeAsO$_{0.85}$F$_{0.15}$ | Au tip | 1.40 | 0.30 | Chen *et. al., Nature* **453**, 1224 (2008) |
| SmFeAsO$_{0.85}$F$_{0.15}$ | Au tip | 1.51 | 0.29 | |
| SmFeAsO$_{0.85}$F$_{0.15}$ | Au tip | 1.66 | 0.19 | |
| SmFeAsO$_{0.8}$F$_{0.2}$ | Ag paint | 1.10 | 0.38 | Gonnelli *et. al, Physica C* **469**, 512 (2009) |
| SmFeAsO$_{0.8}$F$_{0.2}$ | Ag paint | 1.23 | 0.30 | |
| SmFeAsO$_{0.8}$F$_{0.2}$ | Ag paint | 1.41 | 0.25 | Daghero *et. al, Phys. Rev. B* **80**, 060502 (2009) |
| SmFeAsO$_{0.8}$F$_{0.2}$ | Ag paint | 1.12 | 0.31 | |
| SmFeAsO$_{0.9}$F$_{0.1}$ | PtIr or Au tip | 1.02 | 0.85 | Wang *et. al., Supercond. Sci. Technol.* **22**, 015018 (2009) |
| Sn | Ag electrode | 1.44 | 0.38 | Naidyuk *et. al., Physica B* **218**, 122 (1996) |
| SrFe$_{1.74}$Co$_{0.26}$As$_2$ | Au tip | 1.46 | 0.20 | Zhang *et. al., Phys. Rev. B* **82**, 020515 (2010) |
| Ta | Au | 1.23 | 0.60 | Sheet *et. al., Phys. Rev. B* **69**, 134507 (2004) |
| Ta | Au | 1.24 | 0.56 | |
| TbFeAsO$_{0.9}$F$_{0.1}$ | Au tip | 1.27 | 0.40 | |

| Superconductor | Metal contact | Conductance enhancement | Z-value | Reference |
|---|---|---|---|---|
| $TbFeAsO_{0.9}F_{0.1}$ | Au tip | 1.35 | 0.37 | Yates *et. al., New. J. Phys.* **11**, 025015 (2009) |
| $TbFeAsO_{0.9}F_{0.1}$ | Au tip | 1.33 | 0.44 | |
| $TbFeAsO_{0.9}F_{0.1}$ | Au tip | 1.29 | 0.49 | |
| $URu_2Si_2$ | Pt tip | 1.32 | 0.33 | Naidyuk *et. al., EPL* **33**, 557 (1996) |
| $URu_2Si_2$ | Pt tip | 1.07 | 0.43 | |
| $URu_2Si_2$ | Pt tip | 1.10 | 0.33 | |
| $Y_{0.9}Ca_{0.1}Ba_2Cu_3O_7$ | Au tip | 1.13 | 0.68 | Kohen *et. al., Phys. Rev. Lett.* **90**, 207005 (2003) |
| $Y_{0.9}Ca_{0.1}Ba_2Cu_3O_7$ | Au tip | 1.35 | 0.49 | |
| $YBa_2Cu_3O_7$ | Au tip | 1.56 | 0.34 | |
| $YNi_2B_2C$ | Ag tip | 1.22 | 0.43 | Bashlakov *et. al., Supercond. Sci. Technol.* **18**, 1094 (2005) |
| $YNi_2B_2C$ | Cu tip | 1.12 | 0.50 | |
| $YNi_2B_2C$ | Ag tip | 1.12 | 0.63 | Mukhopadhyay *et. al., Phys. Rev. B* **72**, 014545 (2005) |
| $YNi_2B_2C$ | Cu tip | 1.17 | 0.50 | Bashlakov *et. al., J. Low. Temp. Phys.* **147**, 335 (2007) |
| $YNi_2B_2C$ | Cu tip | 1.12 | 0.42 | |
| $YNi_2B_2C$ | Cu tip | 1.13 | 0.50 | |
| $YNi_2B_2C$ | Cu tip | 1.11 | 0.51 | |
| $YNi_2B_2C$ | Cu tip | 1.07 | 0.55 | |
| Zn | Ag | 1.11 | 0.50 | Naidyuk *et. al., Phys. Rev. B* **54**, 16077 (1996) |
| Zn | Ag | 1.09 | 0.51 | Naidyuk *et. al., Phys. Rev. B* **54**, 16077 (1996) |

## 2.4. Comparison of conductance spectra in standard BTK and Dirac-BTK models

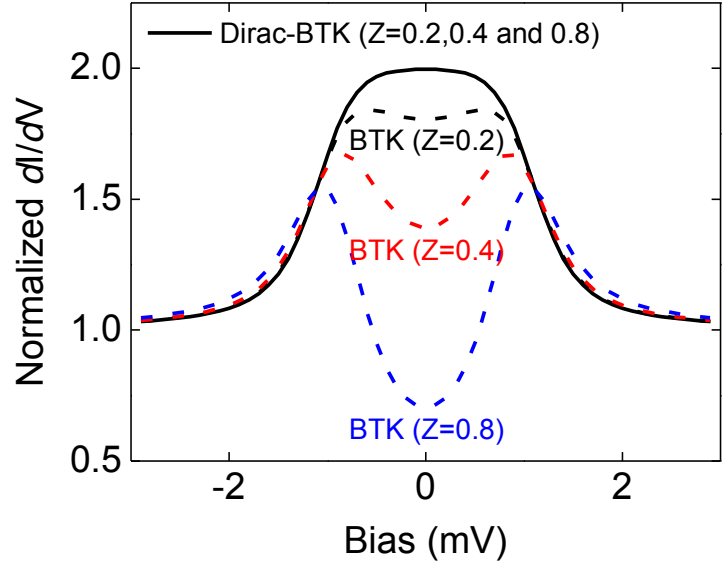

**Figure S6.** Comparison of calculated $dI/dV$ spectra with the standard BTK and the Dirac-BTK models for $Z$=0.2, 0.4, and 0.8 ($\Delta$=1 meV).

Figure S6 shows a comparison of conductance curves according to the standard BTK and the Dirac-BTK models for different $Z$ values, from which it can be clearly seen how conductance spectrum is modified by changing the barrier strength $Z$. In the standard BTK model, the conductance within the superconducting gap gradually decreases with increasing $Z$, whereas the conductance spectra in the Dirac-BTK model remain unchanged regardless of the value of $Z$, as theoretically described in the main text. Such dependency is also captured in the curves in Figure 3 of the main text (*i.e.*, zero-bias normalized $dI/dV$ *vs.* $Z$ according to the standard BTK and the Dirac-BTK models). Figure S7 shows the comparison of the Dirac-BTK and the standard BTK fits to the conductance spectrum of a PtIr-SmB$_6$(20nm)/YB$_6$ contact. When the standard BTK model is used, as expected, the best fit is obtained by setting $Z$ = 0, which then provides an identical fit to the Dirac-BTK (with the same $\Delta$). If we use a more realistic value of $Z$ = 0.39, the standard BTK gives a fit with significant deviation from the experimental curve. As discussed in the main text, this $Z$ = 0.39, is assessed as the realistic value extracted from spectra taken on materials-wise similar heterostructures without complete topological protection, namely, 10 nm SmB$_6$/YB$_6$ and Y-substituted SmB$_6$/YB$_6$ heterostructures. The plot clearly demonstrates that the standard BTK model with a finite and realistic $Z$ cannot reproduce the experimental data showing the perfect Andreev reflection.

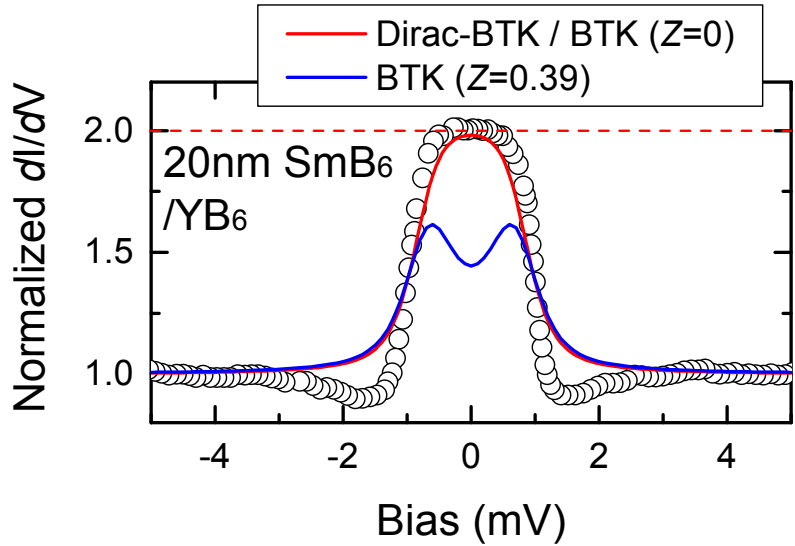

**Figure S7.** Comparison of the Dirac-BTK and the standard BTK fits to the experimental conductance spectrum of a PtIr-SmB$_6$(20nm)/YB$_6$ contact (Fig. 1c in main text). The red curve is the theoretical conductance curve in the Dirac-BTK model and the standard BTK model with $Z = 0$. Both appear identical as expected for the same $\Delta$ (= 0.77). The blue curve is the theoretical standard BTK curve with $\Delta = 0.77$ and $Z = 0.39$ which is assessed from contacts to other heterostructures in this study without perfect Andreev reflection (*i.e.*, the ones with thin SmB$_6$(10 nm) and Y-substituted SmB$_6$ – see main text for details). The effect of nullifying $Z$ due to incorporation of a Dirac material in Andreev reflection process is clearly seen.

## 2.5. Magnetic field-dependent conductance spectra of a point contact with a SmB$_6$/YB$_6$ heterostructure

Applying a magnetic field can break time reversal symmetry (TRS), and the effect can be used as a signature of the perfect Andreev reflection due to Klein tunneling. We have carried out field-dependent conductance spectrum measurements on a device with a thin-film Au-layer for a normal metal (*i.e.*, Au-SmB$_6$/YB$_6$ structure, see Section 1.3 in SI) which provides a stable contact under application of magnetic field, as opposed to a point-contact junction which can potentially suffer from magnetostriction. As shown in Figure S8a, the conductance enhancement is indeed gradually suppressed with increasing magnetic field in both out-of-plane and in-plane field configurations, but the zero-bias conductance decreases more quickly when the magnetic field is applied along the out-of-plane direction compared to when it is applied in-plane (Fig. S8b). On the other hand, the decreasing trend of the superconducting gap ($\Delta$) due to applied field is approximately the same for out-of-plane and in-plane directions (the inset of Fig. S8b). The fact that zero-bias conductance is suppressed more quickly with out-of-plane field thus cannot be explained only by field-induced diminishing of superconductivity in the $SmB_6/YB_6$ heterostructure.

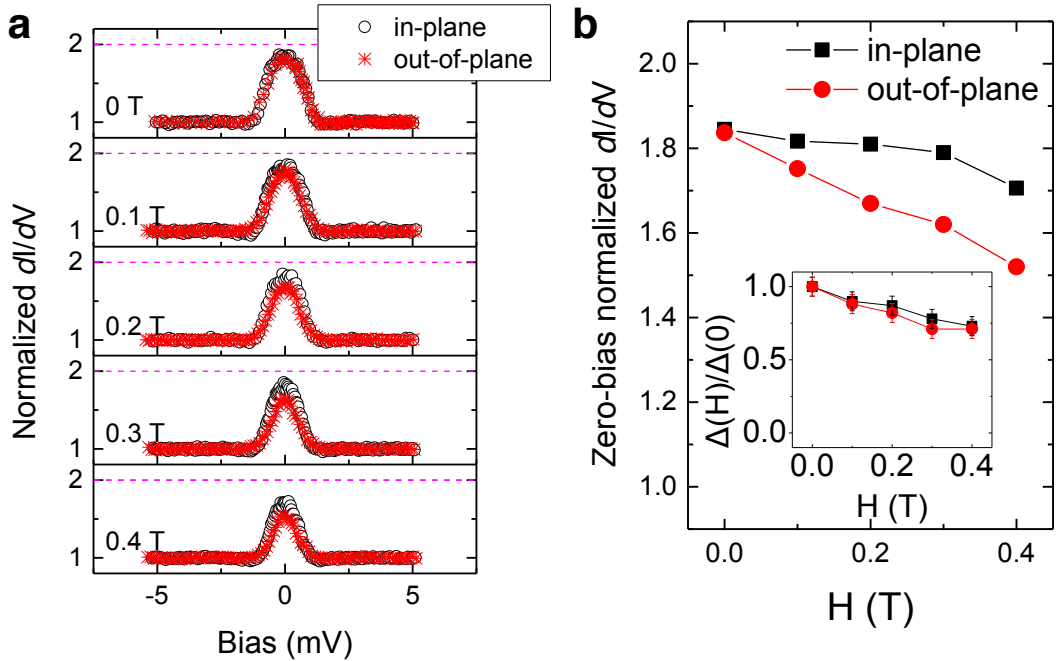

**Figure S8.** (a) Conductance spectra of Au-$SmB_6$/$YB_6$ device under magnetic field applied along in-plane and out-of-plane directions. (b) Zero-bias normalized conductance as a function of magnetic field. The inset shows superconducting order parameter ($\Delta$) as a function of magnetic field normalized by $\Delta$ at 0 T ($\Delta(0)$). $\Delta$ was estimated as the bias voltage point where maximum first derivative of each conductance spectrum occurs under different magnetic fields.

The effect of magnetic field on the helical surface states depends on factors such as the direction of the field, the position of the Fermi level relative to the Dirac point, and the magnitude of the effective $g$-factor. Applying magnetic field parallel to the surface will distort and shift the Dirac cone, but without affecting the spin-momentum locking at the Fermi level[14,15]. However, a magnetic field component perpendicular to the surface will open a gap at the Dirac point and a back-scattering channel by inducing a $z$-component of the electron spins[14-16]. In other words, we expect significant suppression of the conductance when the field is applied out of plane, which is consistent with our observation here.

The observed suppression, however, is not very dramatic in either direction, and we attribute this to the small value of effective $g$-factor of the surface states in SmB$_6$. The size of opened gap or the shift in Fermi surface ($\Delta_B$) due to magnetic field $B$ is proportional to Zeeman energy, $\Delta_B = g_{eff}\mu_B B$, where $g_{eff}$ is the effective $g$-factor of surface states and $\mu_B$ is Bohr magneton[17]. Thus, for small enough $g_{eff}$, the application of $B$ does not weaken topological protection significantly, provided the Fermi level is sufficiently far away from the Dirac point. To the best of our knowledge, the effective $g$-factor for the surface states of SmB$_6$ has not been reported, but the value for the bulk states of SmB$_6$ has been estimated to be $\sim 0.1$[18,19]. It has been reported that the effective $g$-factor of surface states of Bi$_2$Se$_3$ is similar to its bulk value in Bi$_2$Se$_3$ ($g_{eff} \approx 50$)[20]. In the absence of a directly measured value for SmB$_6$ and assuming that its behavior is similar to Bi$_2$Se$_3$, we take the $g$-factor of the surface state of SmB$_6$ to be also $\sim 0.1$.

Recent magnetoresistance studies on SmB$_6$ also suggests a small effective $g$-factor for the surface states of SmB$_6$[19,21,22]. For example, S. Wolgast *et al.*, have reported on very weak field-dependence of the resistance at low temperatures (for instance, $\Delta R/R \sim 2$ % for 80 T at 1.39 K), suggesting that the surface states of SmB$_6$ are extremely robust against applied magnetic field. This is consistent with the gradual suppression of conductance enhancement by magnetic field observed here.

## 2.6. Conductance doubling and conductance dip near the gap

To the best of our knowledge, there have only been two reports in the literature where observed conductance enhancement is larger than 1.9. They are both on Nb-Cu point contacts[23,24] (also see Fig. 3 in the main text and Table S1). The spectra showing conductance doubling therein are reproduced in Fig. S9, and one of our PtIr-SmB$_6$/YB$_6$ spectra is also shown in the figure for comparison. The reported Nb-Cu spectra exhibit distinctive features, namely, conductance dips near the bias voltage corresponding to the superconducting gap energy of Nb (indicated by arrows in Fig. S9). These dips cannot be reproduced using the standard BTK theory solely. Strijkers *et al.* have proposed a model to account for the dips that are intimately tied to the conductance doubling[24]. In this model, when $Z$ is exceptionally small due to a negligible Fermi velocity mismatch as in the special case of Nb-Cu junctions, the interface becomes effectively transparent, which allows the superconducting proximity effect to create a region in the normal metal side with a superconducting order parameter ($\Delta_{prox}$, smaller than the order parameter of the superconductor). In such an instance, the Andreev reflection process is limited to the energy of incident particles within $|\Delta_{prox}|$. According to the model put forth by Strijkers *et al.*, because the quasiparticles in the proximitized layer on the normal metal side can only enter the superconductor side when their energy is outside the energy gap of the superconductor, the conductance spectrum develops large conductance dips near voltages roughly corresponding to the gap energy of the superconductor. Therefore, we attribute the substantial dip feature to $Z \approx 0$ in the case of Nb-Cu junctions. The absence of such feature in our results thus indicates that perfect conductance doubling observed in the PtIr-SmB$_6$/YB$_6$ junctions is of a different origin compared to those in the Nb-Cu contacts. In the case of a contact between PtIr and SmB$_6$, a substantial barrier is expected based just on the significant Fermi velocity mismatch between them (the Fermi velocity of the surface states of SmB$_6$ is $< 10^5$ m/s[2,25,26]). These facts underscore the need for an alternative model to explain the perfect Andreev reflection observed in the PtIr-SmB$_6$/YB$_6$ heterostructures here.

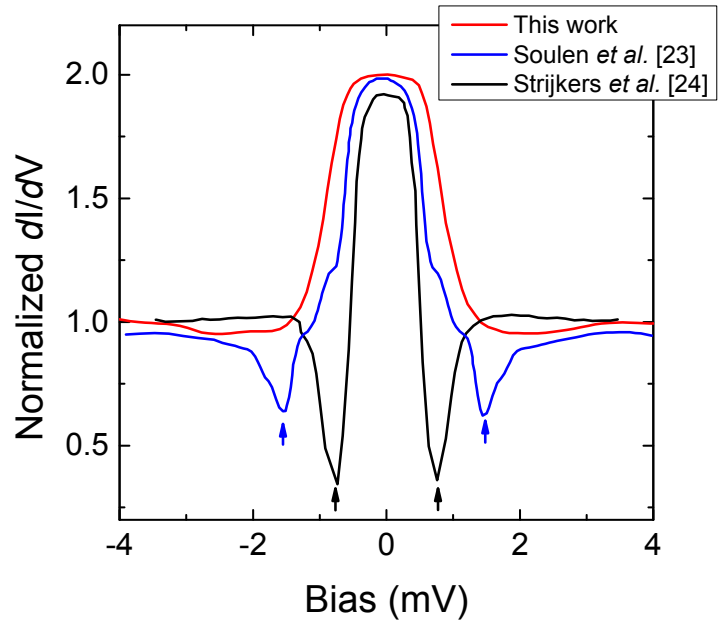

**Figure S9.** Comparison of normalized $dI/dV$ spectrum obtained from the PtIr-SmB$_6$(20 nm)/YB$_6$ junction (this work) with the reported spectra obtained from Nb-Cu junctions[23,24]. The red line is the experimental data obtained from the PtIr-SmB$_6$(20 nm)/YB$_6$ junction in the present study. The black and blue lines are point contact spectra for Nb-Cu junction reported by Soulen *et al.*[23] and Strijker *et al.*[24], respectively. The arrows indicate conductance dips near the $\Delta$. Such dips are not present in our spectrum.

Note that shallow dips observed in the conductance spectra of our PtIr-SmB$_6$/YB$_6$ junctions are common to conductance spectra of various normal metal-superconductor junctions [for example, Refs. 27-31]. They are attributed to the inhomogeneous nature of point contact which can consist of many parallel channels, and excessive current flowing in some of them[28,32].

## 3. Details of the Modified Blonder, Tinkham and Klapwijk theory with the Dirac Hamiltonian

To describe particle transmission and reflection at the interface of a normal metal with topological superconductor, we use the framework developed by Blonder, Tinkham and Klapwijk for normal metal – conventional superconductor boundary (BTK theory)[13]. It describes a plane wave on the normal metal side, which can be either scattered or transmitted to the superconducting side. Since only paired electrons are allowed for energies less than $\Delta$ (for temperatures sufficiently below $T_c$), an electron entering the superconductor has to partner with another electron, thus leaving a positively charged hole moving away from the interface. This is Andreev reflection; it leads to enhancement of the conductance for energies below $\Delta$ (as compared to the normal state conductance). However, this process is very sensitive to various scattering mechanisms, and for an imperfect interface (usually modelled by a repulsive delta-function potential) and significant Fermi velocity mismatch, it is strongly suppressed; the reason is the need to transmit two electrons (vs. a single one for $T > T_c$). For normal metal – conventional s-wave superconductor, these scattering mechanisms can be captured in a single dimensionless parameter $Z$. (In the standard BTK theory, $Z$ is commonly defined as $U_0/\hbar v_F^S$, but modified formula – $Z \equiv \sqrt{(U_0/\hbar v_F^S)^2 + (v_F^N - v_F^S)^2/4 v_F^N v_F^S}$ – can be used to explicitly incorporate the effect of the Fermi velocity mismatch. Here $U_0$ is the potential barrier height, and $v_F^N$ and $v_F^S$ are the Fermi velocities in the normal metal and the superconductor, respectively.) Perfect Andreev reflection implies $Z \approx 0$, and as illustrated in Fig. 3 of the main text, this is very rarely observed in real experiments.

The PtIr-SmB$_6$ interface requires a significant modification of the standard BTK theory. For temperatures below ~50 K, SmB$_6$ is a topological Kondo insulator, with non-trivial gapped bulk and helical surface states[33]. Thus the surface of SmB$_6$ has only half the degrees of freedom of regular electronic system, and in the normal state it is described by a Dirac Hamiltonian, with basis $\Psi = [\psi_{\uparrow,\varepsilon,\boldsymbol{p}}, \psi_{\downarrow,\varepsilon,\boldsymbol{p}}, \psi^*_{\uparrow,-\varepsilon,-\boldsymbol{p}}, \psi^*_{\downarrow,-\varepsilon,-\boldsymbol{p}}]$ as:

$$H_{\mathrm{SmB_6}} = \begin{bmatrix} v_F \boldsymbol{p}.\boldsymbol{\sigma} - \sigma_0\mu & 0 \\ 0 & v_F \boldsymbol{p}.\boldsymbol{\sigma}^* + \sigma_0\mu \end{bmatrix}, \tag{S1}$$

where $\boldsymbol{p}$ is the electron momentum in the $x$-$y$ plane, $\mu$ is the chemical potential, $\sigma_0$, $\sigma_x$, $\sigma_y$, $\sigma_z$ are the identity and the Pauli matrices in the spin space. ARPES measurements have found three Dirac cones at the SmB$_6$ surface[26], but we consider a simplified model with a single Dirac cone. The presence of three Dirac cones in the surface states of SmB₆ may be affected by band folding due to surface reconstruction[26], but the odd number of Dirac cones still ensures the topological nature of the surface states and thus the validity of the Dirac-BTK model[25]. We can easily obtain the eigenstates of this Hamiltonian with both particle and hole components. Since we are typically interested in $|p - p_F| \ll p_F$ ($p_F$ – Fermi momentum) and $\Delta \ll E_F$ ($E_F$ – Fermi energy), we can neglect the small difference in the momenta of the two components.

A normal metal, on the other hand, has a band structure displaying the double degeneracy of the electronic states associated with spin. Note that (although unconventional) wave functions with positive and negative helicities can be used as a basis for describing normal metal states. However, the contact with the TI surface breaks the degeneracy of the two helicities in the normal metal; the wave functions have to match at the boundary, and thus in a region adjacent to the interface only electrons with the helicity permitted on the SmB₆ side are allowed (the effect of the metallic bulk on the surface states is limited by the small effective size of the contact area[34]). The strong spin-orbit coupling of PtIr itself[35,36] can also significantly enhance this process: when PtIr whose bands are split by the large spin-orbit coupling is brought to contact SmB₆, one of the bands can easily be lifted above the Fermi energy due to the topological proximity effect[34,37-41]. We can use projector operators to write this in a formal way: close to the interface the PtIr Hamiltonian becomes $H_{\text{PtIr}} \to P_+ H_{\text{PtIr}} P_+$, where we can write $P_+ = |\hat{\boldsymbol{k}}, +\rangle\langle\hat{\boldsymbol{k}}, +|$, with $\hat{\boldsymbol{k}} = [\cos\theta_k, \sin\theta_k]$ being the unit vector in the direction of propagation ($\theta_k$ is the incidence angle measured from the perpendicular direction to the boundary). It trivially holds that $(P_+)^2 = P_+$ and $P_+ + P_- = 1$, where $P_-$ is the projector on the other helicity. Using both projectors, $H_{\text{PtIr}} = (P_+ + P_-) H_{\text{PtIr}} (P_+ + P_-)$ is identically true in the bulk PtIr. The matching condition at the SmB₆ boundary implies that $P_- = P_0(z)|\hat{\boldsymbol{k}}, -\rangle\langle\hat{\boldsymbol{k}}, -|$, where $P_0(z)$ is a function of the $z$ – the distance away from the SmB₆ surface, with limits $P_0(z \to 0) \to 0$ and $P_0(z \to \text{bulk PtIr}) \to 1$. The interface between a metal and a topological surface has been studied in several recent experimental and theoretical work, which have demonstrated that the states on the nominally normal side indeed inherit some topological properties from TI[34,37-41].

For the experimental setup discussed in the main text, there is additional strong constraint on the allowed transmission processes originating in conservation of momentum and energy. Since the helical states are localized close to the surface and their momenta typically lie along the interface, incoming electrons with finite $p_z$ momenta exhibit total reflection due to the absence of bulk states in $SmB_6$. It is generally not possible for the electrons with $p_z \neq 0$ momenta in the ballistic regime to match the energies and the momenta on both sides of the interface, and they do not contribute to the current. Thus, the transport through the interface is dominated on the PtIr side by electrons with momenta in the plane of the surface states. Since there are ubiquitous microstructural variations at the point-contact boundary, there are many such channels for which this condition is satisfied.

Following the BTK formalism and the discussion above, we model the PtIr-$SmB_6$ boundary as a line dividing normal metal and superconducting regions in the plane of the $SmB_6$ surface states. On both sides, only electrons with the same helicity are allowed. To model a potential barrier at the interface, we add a delta-function potential term $U(x) = U_0 \delta(x)$. The Hamiltonian for the top $SmB_6$ layer can be written in $\Psi = \left[ \psi_{\uparrow,\varepsilon,\boldsymbol{p}}, \psi_{\downarrow,\varepsilon,\boldsymbol{p}}, \psi^*_{\uparrow,-\varepsilon,-\boldsymbol{p}}, \psi^*_{\downarrow,-\varepsilon,-\boldsymbol{p}} \right]$ basis as

$$H_{\text{hetero}} = \begin{bmatrix} v_F \boldsymbol{p}.\boldsymbol{\sigma} - \sigma_0 \mu + \sigma_0 U(x) & i\sigma_y \Delta \\ -i\sigma_y \Delta & v_F \boldsymbol{p}.\boldsymbol{\sigma} + \sigma_0 \mu - \sigma_0 U(x) \end{bmatrix}, \tag{S2}$$

where $\Delta$ is the proximity-induced superconducting gap. Although we assume conventional *s*-wave spin-singlet order parameter in $YB_6$, once it is projected on the low-energy helical states on the surface of $SmB_6$, the pairing term mixes spin-singlet and spin-triplet states (see, for example[42]). Note also that we disregard more exotic possibilities like inter-layer pairing.

To obtain the reflection and transmission coefficients, we need to match at the boundary the plane wave solutions in both regions. However, we have distinct Hamiltonians on the two sides. Moreover, the delta function term coupled with the first order derivatives in the Dirac equation implies a discontinuity of the wave function itself (rather than the more common discontinuity in the first derivatives). Dealing with this problem requires some caution. First, we derive the probability current expression on both sides (since it has to be continuous). We use the general expression $j_x = \sum \Psi^\dagger \hat{v}_x \Psi$, where $\hat{v}_x = \left\{ \frac{\partial \hat{H}}{\partial k_x} \right\}$ and the sum is over spin and momenta. On both sides, this leads to the expression $j_x \sim v_F \tau_z \otimes \sigma_x \sum \Psi^{\dagger} \Psi$ (where $\tau_x$, $\tau_y$, $\tau_z$ are the Pauli matrices in the particle-hole space). Without the delta function term, probability current conservation gives $v_F^S \Psi_S = v_F^N \Psi_N$. To include the effects of the boundary barrier we expand this to $\Psi_S - (v_F^N/v_F^S)\,\Psi_N = -i\,\tau_z \otimes \sigma_x \Psi(0)$. The structure of this term can be obtained by analogy with the delta function potential problem for pure Dirac system - integrating the Hamiltonian in a small vicinity around the interface gives the proper boundary condition. Note that, however, there is an ambiguity in treating the $x = 0$ term. Here "$\Psi(0)$" is given by $\Psi(0) = \frac{1}{2} Z \left( \Psi_S + (v_F^N/v_F^S)\,\Psi_N \right)$, which guarantees the conservation of probability current through the interface. There is additional ambiguity associated with the exact form of $Z$, but two possible choices given in the literature are $Z = (U_0/\hbar v_F^S)$ and $Z = 2 \tan\left( \frac{1}{2} U_0/\hbar v_F^S \right)$ (we use the first one in the main text, but note that they agree in the limit $U_0/\hbar v_F^S \ll 1$)[43].

We are now ready to match the wave functions on both sides. Assuming states with incoming, normal-reflected and Andreev-reflected plane wave components with energy $E$ in the normal part, we need to solve the equation:

$$\frac{v_F^N}{v_F^S}\left\{ (1 - \mathrm{i}\, Z\, \tau_z \otimes \sigma_x) \begin{bmatrix} 1 \\ e^{i\theta_k} \\ 0 \\ 0 \end{bmatrix} + (1 - \mathrm{i}\, Z \tau_z \otimes \sigma_x) r_e \begin{bmatrix} -1 \\ -e^{-i\theta_k} \\ 0 \\ 0 \end{bmatrix} + (1 - \mathrm{i}\, Z\, \tau_z \otimes \sigma_x) r_h \begin{bmatrix} 0 \\ 0 \\ -e^{i\theta_k} \\ 1 \end{bmatrix} \right\} =$$
$$(1 + \mathrm{i}\, Z\, \tau_z \otimes \sigma_x) t_e \begin{bmatrix} u \\ u e^{i\varphi_k} \\ -v e^{\varphi_k} \\ v \end{bmatrix} + (1 + \mathrm{i}\, Z\, \tau_z \otimes \sigma_x) t_h \begin{bmatrix} v \\ -v e^{-i\varphi_k} \\ u e^{-i\varphi_k} \\ u \end{bmatrix}, \quad \text{(S3)}$$

where the transmission angle $\varphi_k$ is fixed by conservation of $y$-component of the momentum: $p_F^N \sin\theta_k = p_F^S \sin\varphi_k$, and $u$ and $v$ are the standard BCS coherence factors: $u^2 = \frac{1}{2}\left( 1 + \frac{\sqrt{E^2 - \Delta^2}}{E} \right)$, $v^2 = \frac{1}{2}\left( 1 - \frac{\sqrt{E^2 - \Delta^2}}{E} \right)$.

The coefficients in front of the plane waves describe: $t_e$ – transmitted electron-like particle, $t_h$ – transmitted hole-like particle, $r_e$ – reflected electron, and $r_h$ – Andreev-reflected hole.

The solutions for $r_e$, $r_h$, $t_e$, and $t_h$ coefficients can be easily obtained from Equation (S3). In particular, the coefficient in front of the normal- and Andreev-reflected states are:

$$r_e = \frac{i\, e^{i\,\theta_k}(u^2 - v^2)((1+Z^2)^2 \sin\theta_k - (1 - 6Z^2 + Z^4 + 4\,i\,Z\,(Z^2 - 1)\cos\theta_k)\sin\varphi_k)}{(u^2 + v^2)(1 + Z^2)^2 \cos\theta_k \cos\varphi_k + (u^2 - v^2)((1+Z^2)^2 - (1 - 6Z^2 + Z^4)\sin\theta_k \sin\varphi_k)} \quad \text{(S4)}$$

(S5)

$$r_h = \frac{2\,u\,v\,(1+Z^2)\cos\theta_k\cos\varphi_k}{(u^2+v^2)(1+Z^2)^2\cos\theta_k\cos\varphi_k + (u^2-v^2)((1+Z^2)^2 - (1-6Z^2+Z^4)\sin\theta_k\sin\varphi_k)}$$

The conductance through the interface is given by

$$G = \frac{dI}{dV} = G_0 \int_{-\chi}^{\chi}(1 - |r_e|^2 + |r_h|^2)\,f_{\theta_k}\cos\theta_k\;d\theta_k, \tag{S6}$$

where $\chi = \arcsin(v_F^S/v_F^N)$ (for $\theta_k > \chi$ there is total reflection of the incoming electrons), $f_{\theta_k}$ models the angular distribution of the incoming electrons, and $G_0$ is a $E$- and $\theta_k$-independent constant.

For finite temperatures we have to include the Fermi distribution function $f(E,T)$. The conductance at a given bias voltage V becomes

$$G(V,T) \sim \int_{-\infty}^{\infty} G(E, T=0)\left(-\frac{\partial f(E-eV,T)}{\partial E}\right)dE \tag{S7}$$

The angular dependence of the normal reflection coefficient goes like $|r_e|^2 \sim g(v_F^N/v_F^S, Z)(\sin\theta_k)^2$ where $g(v_F^N/v_F^S, Z)$ is a complicated algebraic function with the limit $g(v_F^N/v_F^S \to 1, Z \to 0) \to g_1(1 - v_F^N/v_F^S)^2 + g_2 Z^2$ ($g_1$ and $g_2$ as numerical coefficients). Note that in this limit the effects of the boundary barrier and the Fermi velocities mismatch are additive. In general, in the modified BTK theory (*i.e.*, the Dirac-BTK) there are mixed $v_F^N/v_F^S$ and $Z$ terms in the expression for $r_e$. This is in contrast with the standard BTK theory, in which the effects of $v_F^N/v_F^S$ can be fully absorbed in a renormalized $Z$ for arbitrary Fermi velocity mismatch. It is important to point out that reflection at $\theta_k = 0$ as a normal particle requires a complete spin-flip, which is forbidden by time-reversal symmetry (*i.e.*, the overlap of the two spin states is zero) and thus $r_e(\theta_k = 0) = 0$ (Fig. S10a). However, electron reflection is not completely forbidden for electrons with oblique incident angles. As depicted in Fig. S10b, the normal reflection of such particle is now allowed even under the presence of spin-momentum locking, since the spin of the incident and reflected electrons are not completely opposite to each other.

As a model of the angular distribution of the incoming electrons $f_{\theta_k}$ we choose a Gaussian function:

$$f_{\theta_k} = e^{-\theta_k^2/\varrho_\vartheta^2} \tag{S8}$$

where $\varrho_\theta$ parametrizes the spread of the distribution (*i.e.*, full width at half maximum of the Gaussian function). The limits $\varrho_\theta \gg \pi$ and $\varrho_\theta \ll \pi$ describe a uniform distribution and a very narrow distribution, respectively. Figure S10c shows variation in the normal reflection and Andreev reflection probabilities as a function of the incident angle. Note then that this means that in the quasi-one-dimensional case with $\varrho_\theta \ll \pi$, where only $\theta_k \approx 0$ channels contribute to the transport, there is perfect transmission irrespective of the height of the barrier strength and the Fermi velocity mismatch. As shown in Fig. S10d, the zero-bias conductance decreases with increasing $\varrho_\theta$ and only a narrow angular distribution (*i.e.*, quasi-one-dimensional transport) can reproduce the perfect conductance doubling observed in the conductance spectra of PtIr-SmB$_6$/YB$_6$. For energies below the gap (*i.e.*, $eV < \Delta$) we have $r_h(\theta_k = 0) = 1$ (a direct consequence of $r_e = 0$). In the opposite case, for $eV \gg \Delta$, we have $r_h \to 0$ (due to $v \to 0$). Combining these two results with Eq. S6 leads to the observed conductance doubling: $G(|eV| < \Delta)/G(|eV| \gg \Delta) = 2$.

The conductance doubling observed here indicates occurrence of one-dimensional channels in our contact. The presence of particular nanostructures at the surface of SmB$_6$ can modify the effective distributions from a uniform $f_{\theta_k}$ to a narrow one. The formation of quasi-one-dimensional contact here is attributed to reconstructed surface atomic structures (*i.e.*, Sm 2×1 surface) on the surface of SmB$_6$ revealed by recent scanning tunneling microscopy (STM)[44,45], angle-resolved photoemission spectroscopy (ARPES)[25,26] and low-energy electron diffraction (LEED)[46]. It is the instability of the ideal Sm 1×1 or B 1×1 polar surface configuration, which leads to the surface reconstruction. C. E. Matt *et al.* have reported on the formation energies of different surface configurations, where Sm 2×1 and B $\sqrt{2}$×$\sqrt{2}$ surface configurations are found to have much lower formation energies than other configurations[47]. Both Sm 2×1 and B $\sqrt{2}$×$\sqrt{2}$ surface configurations would provide parallel quasi-one-dimensional channels at the point contact.

It is also possible that the small effective size of the tip is playing a role in the quasi-one-dimensional transport. Namely, it can lead a finite-size quantization of the electronic states and separation of the different angular momentum channels (for approximately axially-symmetric tip). The lowest-energy $l = 0$ channel is then equivalent to the normal incidence for the infinite plane geometry considered here.

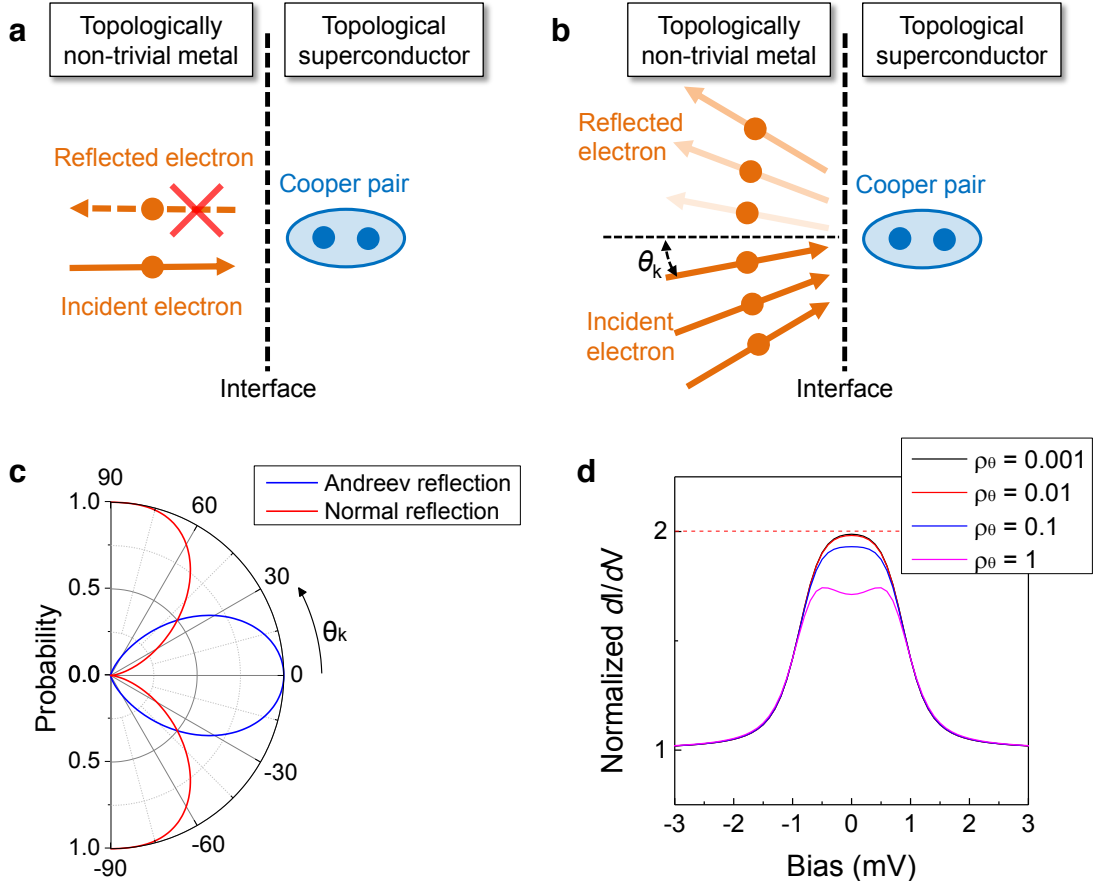

**Figure S10.** Schematic of electron incident and reflection processes in Andreev reflection at the interface of a topologically non-trivial metal and a topological superconductor for (a) normal incidence and (b) oblique incidence cases where the reflection probability is depicted by shade, *i.e.*, lighter color indicates smaller reflection probability. In the case of normal incidence, the electron reflection is not allowed (as indicated by X). Since the spin of obliquely reflected electron is not completely opposite to the spin determined by its momentum, the normal reflection is not perfectly prohibited. (c) Polar plot of the normal reflection (red) and Andreev reflection (blue) probabilities in the Dirac-BTK theory. (d) Simulated conductance spectra for different $\varrho_\theta$ (*i.e.*, full width at half maximum of Gaussian angle distribution = 0.001, 0.01, 0.1, and 1) in the Dirac-BTK model. With increasing angle distribution of incident particles, the zero-bias conductance decreases.

In the more general case, with all possible angles of incident up to the critical $\chi$ included, the conductance enhancement does not lead to perfect doubling (Fig. S10d). Nevertheless, the presence of the highly transmitting channels close to $\theta_k \approx 0$ reduces the effects of barrier at the boundary and the Fermi velocities mismatch. Even for $v_F^N/v_F^S$ and $Z$ combinations for which the interface would be in a tunnel regime according to the standard BTK theory, the Dirac-BTK model calculation shows sizable conductance enhancement.